\documentclass[useAMS,usenatbib]{mn2e}
\usepackage[colorlinks,citecolor=blue]{hyperref}
\usepackage{textcomp}
\usepackage{subfigure}
\usepackage{verbatim}
\usepackage{bm}
\usepackage{hyperref}
\usepackage{amsmath}
\usepackage{graphicx}
\usepackage{epstopdf}
\usepackage{amssymb}
\usepackage{extarrows}
\usepackage{color}
\usepackage{CJK}
\usepackage{cancel}
\usepackage{ulem}
\usepackage[utf8]{inputenc}
\newcommand{\ba}{\begin{eqnarray}}
\newcommand{\ea}{\end{eqnarray}}
\newcommand{\be}{\begin{equation}}
\newcommand{\ee}{\end{equation}}
\newcommand{\gr}{\mathrm{GR}}

\newcommand{\au}{\mathrm{AU}}

\newcommand{\IN}{\mathrm{in}}

\newcommand{\OUT}{\mathrm{out}}

\newcommand{\eff}{\mathrm{eff}}

\def\e1{e_1^2}

\title[Evolution of Stellar Orbits Around Merging BHB]
{Evolution of Stellar Orbits Around Merging Massive Black-Hole Binary}
\author[Liu, \& Lai]
{Bin Liu$^{1}$, Dong Lai$^{2}$\\
$^{1}$ Niels Bohr International Academy,
Niels Bohr Institute, Blegdamsvej 17, 2100 Copenhagen, Denmark\\
$^{2}$ Cornell Center for Astrophysics and Planetary Science,
Department of Astronomy, Cornell University, Ithaca, NY 14853, USA\\
}
\date{November 2021}

\begin{document}
\maketitle

\begin{abstract}
{We study the long-term orbital evolution of stars around a merging
massive or supermassive black-hole (BH) binary, taking into account
the general relativistic effect induced by the BH spin.  When the BH
spin is significant compared to and misaligned with the binary orbital
angular momentum, the orbital axis ($\hat{\bm{l}}$) of the
circumbinary star can undergo significant evolution during the binary orbital
decay driven by gravitational radiation. Including
the spin effect of the primary (more massive) BH, we find that
starting from nearly coplanar orbital orientations, the orbital axes
$\hat{\bm{l}}$ of circumbinary stars preferentially evolve towards the
spin direction after the merger of the BH binary, regardless of the
initial BH spin orientation.  Such alignment phenomenon, i.e., small
final misalignment angle between $\hat{\bm{l}}$ and the spin axis of
the remanent BH $\hat{\bm{S}}$, can be understood analytically using
the principle of adiabatic invariance. For the BH binaries with
extremely mass ratio ($m_2/m_1\lesssim0.01$), $\hat{\bm{l}}$ may
experience more complicated evolution as adiabatic invariance breaks
down, but the trend of alignment still works reasonably well when the
initial binary spin-orbit angle is relatively small. Our result
suggests that the correlation between the orientations of stellar
orbits and the spin axis of the central BH could provide a potential
signature of the merger history of the massive BH.
}
\end{abstract}


\section{Introduction}
\label{sec 1}

Massive black-hole (BH) binaries, with orbital separations $\lesssim
10$~pc, are natural products of galaxy mergers
\citep[e.g.,][]{Begelman 1980,Milosavljevic 2001,Milosavljevic 2005,Escala 2005,
Mayer 2007,Dotti 2007,Cuadra 2009,Chapon 2013,Fragione IMBH}. Significant observational efforts have been devoted to
searching for such binaries, and a number of candidate systems have
been detected using various techniques \citep[e.g.,][]{Sillanpaa 1988,
Komossa 2003,Komossa 2008,Rodriguez 2006,Bianchi 2008,
Bogdanovic 2009,Boroson 2009,Dotti 2009,Comerford 2009,Green 2010,Deane 2014,Liu X 2014,Bansal 2017,
Comerford 2018,De Rosa 2019}.
These massive BH binaries (BHBs) are likely surrounded by stars (or compact
objects) associated with the merging galaxies. Alternatively, the stars could form
in a circumbinary disk or be captured by the disk from a nuclear star cluster \citep[e.g.,][]{Tagawa 2020,Tagawa 2021}.
For sufficiently small orbit separations, the massive binary BHs experience
orbital decay and eventually merge, producing low-frequency gravitational waves (GWs).
How would the orbits of the circumbinary stars change?

The secular gravitational interaction between a central binary and a
surrounding object dictates the long-term evolution the system.
For a hierarchical triple (with the semi-major axis $a_{\rm out}$ of
the outer orbit much larger than that of the inner orbit $a_{\rm in}$),
the secular evolution equations for arbitrary orbital
eccentricities and orientations can be derived using expansion in
$a_{\rm in}/a_{\rm out}$ [see \citet{Ford} for the equations to
the octupole order; more compact equations in the vector form can be
found in \citet{Liu et al 2015,Petrovich 2015}]. Such systems may exhibit
excitations/oscillations in eccentricities and inclinations in both
the inner and outer orbits
\citep[e.g. the well-known Lidov-Kozai effect;][]{von Zeipel,Lidov,Kozai,Naoz 2016}.
In general, the evolution
can be highly irregular when the octupole effects are significant.
If the outer body has a negligible mass
compared to the inner binary, the dynamics of the outer body becomes simpler
and analytical results can be obtained \citep[e.g.,][]{Farago,Zhou JiLin}.
In particular, the inner eccentric binary can drive significant inclination evolution
of the outer orbit \citep[e.g.,][]{JJ 2018} and
produce orbit flipping from extreme eccentricity excitation \citep[e.g.,][]{Naoz 2017}.
\citet{Vinson 2018} carried out a systematic study of the (secular) restricted three-body problem
by expanding the potential to the hexadecapolar order
\citep[see also][]{Gallardo 2012} and identified various secular resonances.

In this paper, we study the secular evolution of stellar orbits around
an inner massive BHB undergoing GW-induced orbital decay. We are
particularly interested in the case of inner massive BHBs with
relatively small mass ratios, such that the spin of the primary
BH may play an important role.
To the Newtonian leading order,
the (inner) massive BHB makes the (outer) stellar orbit precess around the inner binary.
However, when the BH spin is significant compared to the (inner) binary orbital
angular momentum, the inner orbit axis
undergoes Lens-Thirring (LT) precession around the BH spin axis.
Therefore, the angular momentum axis of the stellar orbit
can also be affected by the LT precession in an indirective way.
In several recent studies \citep[][]{Liu SMBH,Liu SMBH 2,Liu SMBH 3}, we have
shown that the GR effects induced by a spinning tertiary SMBH plays an
important role in the evolution of an inner stellar-mass binary.
Here, we extend our previous studies to the ``inverse" secular problem, in which
the tertiary is essentially a test mass.
By evolving the inner massive BHB until merger, we seek to
identify the correlation (or signature) between the distribution of
the surrounding stellar orbits and the final spin orientation of the BHB
merger remanent.

This paper is organized as follows.
In Section \ref{sec 2}, we review the essential GR effects
in the ``BHB$+$outer test particle" system
and present the secular equations in Post-Newtonian (PN) theory.
In section \ref{sec 3}, we identify different dynamical behaviors of the outer orbit
for different parameters of the system.
We perform analytical calculations of the final spin-orbit misalignment angles
using the principle of adiabatic invariance.
In Sections \ref{sec 4} and \ref{sec 5},
we explore the final configurations of the stellar orbits at different distances
from the central BHB, considering a range of mass ratios of BHB,
coplanar/inclined initial orientations and eccentricities of the stellar orbits.
We summarize our main results in Section \ref{sec 6}.

\section{Evolution Equations}
\label{sec 2}

\begin{figure}
\begin{centering}
\includegraphics[width=8.5cm]{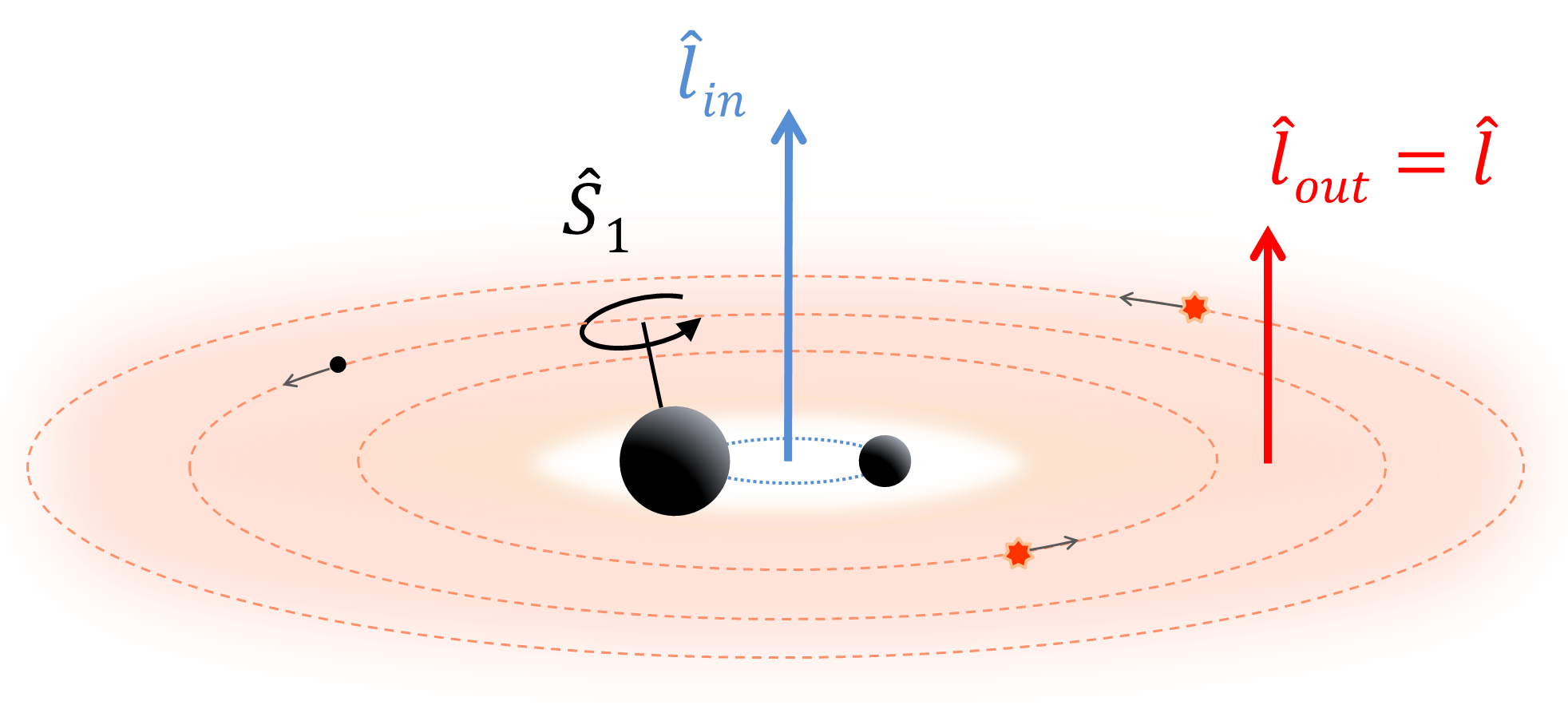}
\caption{Schematic view of the set-up.
We consider a massive BH binary at the center and a stellar disk around the binary.
The inner orbit corresponds to the massive BH binary and
has the unit vector of angular momentum $\hat{\bm{l}}_\IN$.
The primary component of the BH binary is assumed to be fast rotating
with a spin unit vector ($\hat{\bm{S}}_1$).
The outer stellar orbit is the test-particle orbit,
which has the angular momentum unit vector $\hat{\bm{l}}_\OUT=\hat{\bm{l}}$.
}
\label{fig:Configuration}
\end{centering}
\end{figure}

We first review the secular dynamics of massless particles around a massive binary.
Consider a black-hole binary (BHB) with semimajor axis $a_\IN$, eccentricity vector $\bm {e}_\IN$,
total mass $m_{12}\equiv m_1+m_2$ (where $m_1$ and $m_2$ are the individual masses)
and reduced mass $\mu_\IN\equiv m_1m_2/m_{12}$.
The outer test particle moves around the BHB with semimajor axis $a_\OUT$, eccentricity $\bm{e}_\OUT$.
The orbital angular momenta of two orbits are
$\bm{L}_\IN\equiv L_\IN \hat{\bm{l}}_\IN=\mu_\IN\sqrt{G m_{12}a_\IN(1-e_\IN^2)}\hat{\bm{l}}_\IN$ and
$\bm{L}_\OUT\equiv L_\OUT \hat{\bm{l}}_\OUT$ (see Figure \ref{fig:Configuration}).
Throughout the paper, for convenience of notation, we will frequently omit the subscript ``$\OUT$"
for the outer orbit.
The evolution of the system is governed by the double-averaged (DA; averaging over both the inner and outer orbital periods)
secular equations of motion.

For the inner binary, we set $e_\IN=0$.
The primary BH ($m_1$) in the binary has spin $\bm{S}_1=S_1\hat{\bm{S}}_1=(\chi_1G m_1^2/c)\hat{\bm{S}}_1$,
where $\chi_1\leqslant1$ is the Kerr parameter.
Throughout the paper, we assume $S_2\ll S_1$, thus neglecting the dynamical effect of the spin of the low-mass secondary ($m_2$);
this approximation allows some of the dynamical spin-orbit behaviors to be understood analytically (see Section \ref{sec 3}).
However, all the equations showed below are valid for arbitrary mass ratio of the inner binary.
The angular momentum $\bm{L}_\IN$ evolves according to
\be
\frac{d \bm{L}_\IN}{dt}=\frac{d \bm{L}_\IN}{dt}\bigg|_{\mathrm{GW}}
+\frac{d \bm{L}_\IN}{dt}\bigg|_{\mathrm{L_\IN S}}~,\label{eq:Full Kozai 1}
\ee
where the two terms represent dissipation due to gravitational waves (GW) emission and the spin-orbit coupling, respectively.
Gravitational radiation draws energy and angular momentum from the BH orbit,
with \citep[e.g.,][]{Peters 1964}
\be
\frac{d \bm{L}_\IN}{dt}\bigg|_{\mathrm{GW}}=
-\frac{32}{5}\frac{G^3}{c^5}\frac{\mu_\IN m_{12}^2}{a_\IN^4}\bm{L}_\IN.\label{eq:GW 1}
\ee
For reference, the merger time due to GW radiation of a binary with the
initial semi-major axis $a_\IN$ is given by
\ba\label{eq:Tmerger}
T_\mathrm{m}&&=\frac{5c^5 a_\IN^4}{256 G^3 m_{12}^3}\frac{(1+q)^2}{q}\\
&&\simeq10^{10}\bigg(\frac{10^6M_\odot}{m_{12}}\bigg)^3\bigg(\frac{(1+q)^2/q}{12}\bigg)
\bigg(\frac{a_\IN}{224\au}\bigg)^4\mathrm{yrs}\nonumber,
\ea
where we have introduced the mass ratio $q\equiv m_2/m_1$.

Spin-orbit coupling (1.5 PN effect) induces mutual precession of $\hat{\bm{L}}_\IN$ around $\hat{\bm{S}}_1$
 \citep[e.g.,][]{Barker 1975}:
\be
\frac{d \bm{L}_\IN}{dt}\bigg|_{\mathrm{L_\IN S}}=
\Omega_\mathrm{L_\IN S}\hat{\bm{S}}_1\times\bm{L}_\IN,\label{eq:GR LS L}
\ee
where
\be\label{eq:Omega LS}
\Omega_\mathrm{L_\IN S}=\frac{G S_{1}(4+3m_{2}/m_{1})}{2c^2a_\IN^3}.
\ee
The spin vector $\bm{S}_1$ follows
\be\label{eq:spin}
\frac{d \bm{S}_1}{dt}=\Omega_\mathrm{SL_\IN}\hat{\bm{l}}_\IN \times \bm{S}_1,
\ee
with
\be\label{eq:desitter rate}
~~~\Omega_\mathrm{SL_\IN}=\Omega_\mathrm{L_\IN S}\frac{L_\IN}{S_1}=
\frac{3 G n_\IN (m_{2}+\mu_\IN/3)}{2 c^2 a_\IN},
\ee
where $n_\IN=(G m_{12}/a_\IN^3)^{1/2}$ is the mean motion of the inner binary.

The time evolution equations of the outer orbital angular momentum axis $\hat{\bm{l}}$
and eccentricity $\bm{e}$ vectors are given by
\begin{eqnarray}
&&\frac{d \hat{\bm{l}}}{dt}=\frac{d \hat{\bm{l}}}{dt}\bigg|_\mathrm{L_\OUT L_\IN}^\mathrm{(N)}
+\frac{d \hat{\bm{l}}}{dt}\bigg|_\mathrm{L_\OUT L_\IN}^\mathrm{(GR)}
+\frac{d \hat{\bm{l}}}{dt}\bigg|_\mathrm{L_\OUT S}~,\label{eq:Full Kozai 3}\\
&&\frac{d \bm{e}}{dt}=\frac{d \bm{e}}{dt}\bigg|_\mathrm{L_\OUT L_\IN}^\mathrm{(N)}
+\frac{d \bm{e}}{dt}\bigg|_\mathrm{L_\OUT L_\IN}^\mathrm{(GR)}
+\frac{d \bm{e}}{dt}\bigg|_{\mathrm{GR}}
+\frac{d \bm{e}}{dt}\bigg|_\mathrm{L_\OUT S}~.\label{eq:Full Kozai 4}
\end{eqnarray}

The precession of $\hat{\bm{l}}$ around $\hat{\bm{l}}_\IN$ includes the Newtonian and GR components.
The Newtonian precession can be described in the quadruple order
\be
\frac{d \hat{\bm{l}}}{dt}\bigg|_\mathrm{L_\OUT L_\IN}^\mathrm{(N)}
=-\Omega_\OUT^\mathrm{(N)}
(\hat{\bm{l}}_\IN\cdot\hat{\bm{l}})\hat{\bm{l}}_\IN\times\hat{\bm{l}},\label{eq:LinLout Lout N}
\ee
with
\be\label{eq: Omega OUT Newtonian}
\Omega_\OUT^\mathrm{(N)}=\frac{3}{4}\frac{\mu_\IN a_\IN^2}{m_{12}a^2}\frac{n}{(1-e^2)^2},
\ee
where $n=(Gm_{12}/a^3)^{1/2}$.
Note that since $e_\IN=0$, the high order Newtonian perturbation acting on the
outer orbit can be ignored.
Similarly,
\be
\frac{d \bm{e}}{dt}\bigg|_\mathrm{L_\OUT L_\IN}^\mathrm{(N)}
=-\Omega_\OUT^\mathrm{(N)}
\bigg\{(\hat{\bm{l}}_\IN\cdot\hat{\bm{l}})\hat{\bm{l}}_\IN\times\bm{e}
-\Big[\frac{1}{2}-\frac{5}{2}(\hat{\bm{l}}_\IN\cdot\hat{\bm{l}})^2\Big]
\hat{\bm{l}}\times\bm{e}\bigg\}.
\label{eq:LinLout eout N}
\ee

The GR components are given by \citep[e.g.,][]{Liu SMBH, Liu SMBH 2}
\ba
\frac{d \hat{\bm{l}}}{dt}\bigg|_\mathrm{L_\OUT L_\IN}^{(\gr)}=&&\Omega_\OUT^{(\gr)}
\hat{\bm{l}}_\IN\times\hat{\bm{l}},\label{eq:LinLout Lout GR}\\
\frac{d \bm{e}}{dt}\bigg|_\mathrm{L_\OUT L_\IN}^{(\gr)}=&&\Omega_\OUT^{(\gr)}\hat{\bm{l}}_\IN\times\bm{e}
-3\Omega_\OUT^{(\gr)}(\hat{\bm{l}}_\IN \cdot \hat{\bm{l}})\hat {\bm{l}}\times\bm{e},\label{eq:LinLout eout GR}
\ea
with
\be\label{eq:GR LinLout rate}
\Omega_\OUT^{(\gr)}=\frac{2G\mu_\IN n}{c^2}\sqrt{\frac{a_\IN}{a^3(1-e^2)^3}}.
\ee

GR (1-PN correction) introduces pericenter precession of the outer binary,
\be\label{eq:e GR outer}
\frac{d \bm{e}}{dt}\bigg|_{\mathrm{GR}}=\Omega_\mathrm{GR,out}\hat{\bm{l}}\times\bm{e},
\ee
with
\be\label{eq:GR outer}
\Omega_\mathrm{GR,out}=3n\frac{Gm_{12}}{ac^2(1-e^2)}.
\ee

Finally, the spin-orbit coupling also induces the precession of $\hat{\bm{l}}$ around $\bm{S}_1$:
\ba
\frac{d \hat{\bm{l}}}{dt}\bigg|_{\mathrm{L_\OUT S}}&&=
\Omega_\mathrm{L_\OUT S}\hat{\bm{S}}_1\times\hat{\bm{l}},\label{eq:LoutS Lout}\\
\frac{d \bm{e}}{dt}\bigg|_{\mathrm{L_\OUT S}}
&&=\Omega_\mathrm{L_\OUT S}\hat{\bm{S}}_1\times\bm{e}-3\Omega_\mathrm{L_\OUT S}
\big(\hat{\bm{l}}\cdot\hat{\bm{S}}_1\big)\hat{\bm{l}}\times\bm{e},
\label{eq:LoutS eout}
\ea
where
\be\label{eq:Omega LoutS}
\Omega_\mathrm{L_\OUT S}=\frac{2G S_{1}}{c^2a^3(1-e^2)^{3/2}}.
\ee
By comparing Equations (\ref{eq:GR LinLout rate}) and (\ref{eq:Omega LoutS}), we find that
$\Omega_\mathrm{L_\OUT S}/\Omega_\OUT^{(\gr)}=S_1/L_\IN$(at $e_\IN=0$).

\section{Analytical Results}
\label{sec 3}

\subsection{Different types of $\hat{\bm{l}}$ behaviors}
\label{sec 3 1}

To develop an analytic understanding of the dynamics,
we assume the outer test particle has a circular orbit.
If we define $\bm{J}\equiv J\hat{\bm{J}}=\bm{L}_\IN+\bm{S}_1$, Equation (\ref{eq:GR LS L}) gives
\be\label{eq:GR LJ}
\frac{d \hat{\bm{l}}_\IN}{dt}\bigg|_{\mathrm{L_\IN S}}
=\frac{d \hat{\bm{l}}_\IN}{dt}\bigg|_{\mathrm{L_\IN J}}
=\Omega_\IN\hat{\bm{J}}\times\hat{\bm{l}}_\IN,
\ee
with
\be\label{eq:Omega LJ}
\Omega_\IN=\Omega_{\mathrm{L_\IN S}}\frac{J}{S_1}=\frac{G J(4+3m_{2}/m_{1})}{2c^2a_\IN^3}.
\ee
Combining Equations (\ref{eq:LinLout Lout N}), (\ref{eq:LinLout Lout GR}) and (\ref{eq:LoutS Lout}),
we find that the orbital axis $\hat{\bm{l}}$ of the test particle evolves according to
\be\label{eq:GR LOUT}
\frac{d \hat{\bm{l}}}{dt}=-\Omega_\OUT^\mathrm{(N)}(\hat{\bm{l}}_\IN\cdot
\hat{\bm{l}})\hat{\bm{l}}_\IN\times\bm{l}
+\Omega_\OUT^{'(\gr)}\hat{\bm{J}}\times\hat{\bm{l}},
\ee
where
\be
\Omega_\OUT^{'(\gr)}=\frac{\Omega_\OUT^{(\gr)}J}{L_\IN}.
\ee

In the absence of GW dissipation, $\hat {\bm{l}}_\IN$ rotates around $\hat{\bm{J}}$
at a constant rate, $\Omega_\IN$,
so it is useful to consider the evolution of $\hat {\bm{l}}$
in the frame corotating with $\hat {\bm{l}}_\IN$.
Combining Equations (\ref{eq:GR LJ}) and (\ref{eq:GR LOUT}), we have
\be\label{eq:LOUT rot}
\bigg(\frac{d\hat{\bm{l}}}{dt}\bigg)_\mathrm{rot}=
\bigg[-\Omega_\OUT^\mathrm{(N)}(\hat {\bm{l}}_\IN\cdot \hat {\bm{l}})\hat {\bm{l}}_\IN
+(\Omega_\OUT^{'(\gr)}-\Omega_\IN)\hat{\bm{J}}\bigg]\times\hat{\bm{l}}.
\ee
The corresponding Hamiltonian can be given by
\be
\mathcal{H}=-\frac{1}{2}\Omega_\OUT^{(\mathrm{N})}(\hat {\bm{l}}_\IN\cdot\hat {\bm{l}})^2
+(\Omega_\OUT^{'(\gr)}-\Omega_\IN)(\hat {\bm{J}}\cdot\hat {\bm{l}}).
\ee
We define the dimensionless Hamiltonian
\be\label{eq: Hamiltonian dimensionless}
\bar{\mathcal{H}}=\frac{\mathcal{H}}{\Omega_\OUT^{(\mathrm{N})}}
=-\frac{1}{2}(\hat {\bm{l}}_\IN\cdot\hat {\bm{l}})^2
+\bigg(\lambda\frac{J}{L_\IN}-\eta\bigg)(\hat {\bm{J}}\cdot\hat {\bm{l}}),
\ee
where we have introduced the dimensionless ratios
\be\label{eq: eta}
\lambda=\frac{\Omega_\OUT^{'(\gr)}}{\Omega_\OUT^{(\mathrm{N})}}, ~~~
\eta=\frac{\Omega_\IN}{\Omega_\OUT^{(\mathrm{N})}}.
\ee

\begin{figure}
\begin{centering}
\includegraphics[width=8.5cm]{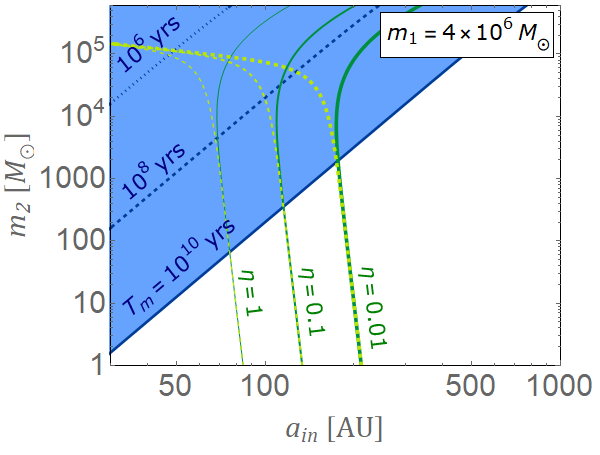}
\caption{Parameter space in the $m_2$-$a_\IN$ plane.
The primary BH has mass $m_1=4\times10^6 M_\odot$,
and the orbits are assumed to be circular.
The blue region corresponds to the BHB that can merge
within $10^{10}$yrs, and the
dark blue solid, dashed, dotted lines are evaluated at
$T_\mathrm{m}=10^{10}$, $10^{8}$ and $10^{6}$ yrs (Equation \ref{eq:Tmerger}), respectively.
The Green lines show different values of $\eta$ (see Equation \ref{eq: eta})
evaluated for $a=a_\OUT=a_{\OUT,\mathrm{c}}$ (see Equation \ref{eq: aoutc}).
The solid/dashed lines are the results of
 $\theta_{\mathrm{S},\IN}=0^\circ, 180^\circ$.
}
\label{fig:parameter space}
\end{centering}
\end{figure}

Note that compared to the Newtonian precession $(\Omega_\OUT^{(\mathrm{N})})$,
the GR precession $(\Omega_\OUT^{'(\gr)})$ of $\hat {\bm{l}}$ is only important near the merger of the inner binary.
We thus ignore the $\lambda$ term in our analytical analysis.
Depending on the value of $\eta$, we expect three possible $\hat {\bm{l}}$ behaviors:
(i) For $\eta\ll1$, $\hat {\bm{l}}$ closely follows $\hat {\bm{l}}_\IN$,
maintaining an approximately constant $I=\cos^{-1}(\hat {\bm{l}}_\IN\cdot\hat {\bm{l}})$.
(ii) For $\eta\gg1$, $\hat {\bm{l}}$ effectively precesses around $\hat {\bm{J}}$
with approximately constant $\theta=\cos^{-1}(\hat {\bm{l}}\cdot\hat {\bm{J}})$.
(iii) When $\eta\sim1$,
a resonance behavior of $\hat {\bm{l}}$ may occur,
and large oscillation in $I$ can be generated.

Figure \ref{fig:parameter space} presents the parameter space indicating
the how the dynamical behavior of $\hat {\bm{l}}$ can change
during the merger of the inner BHB.
We set the primary component of the BHB to be $m_1=4\times10^6 M_\odot$,
and vary the mass of the secondary component ($m_2$)
and the semimajor axis of the BHB ($a_\IN$).
The contours of constant $\eta$ are evaluated for the closest stable test particle orbits
around the binary \citep[][]{Holman 1999}:
\ba\label{eq: aoutc}
&&a_{\OUT,\mathrm{c}}=(1.6+5.1e-2.22e^2+4.12\mu_\mathrm{c}-4.27e\mu_\mathrm{c}\nonumber\\
&&~~~~~~~~~~~-5.09\mu_\mathrm{c}^2+4.61e^2\mu_\mathrm{c}^2)a_\IN,
\ea
where $\mu_\mathrm{c}=m_2/m_{12}$.
We see that for a given $m_2$, as $a_\IN$ decreases, $\eta$ increases
and the outer orbit may experience three types dynamical behaviors successively.

\begin{figure*}
\begin{centering}
\includegraphics[width=16cm]{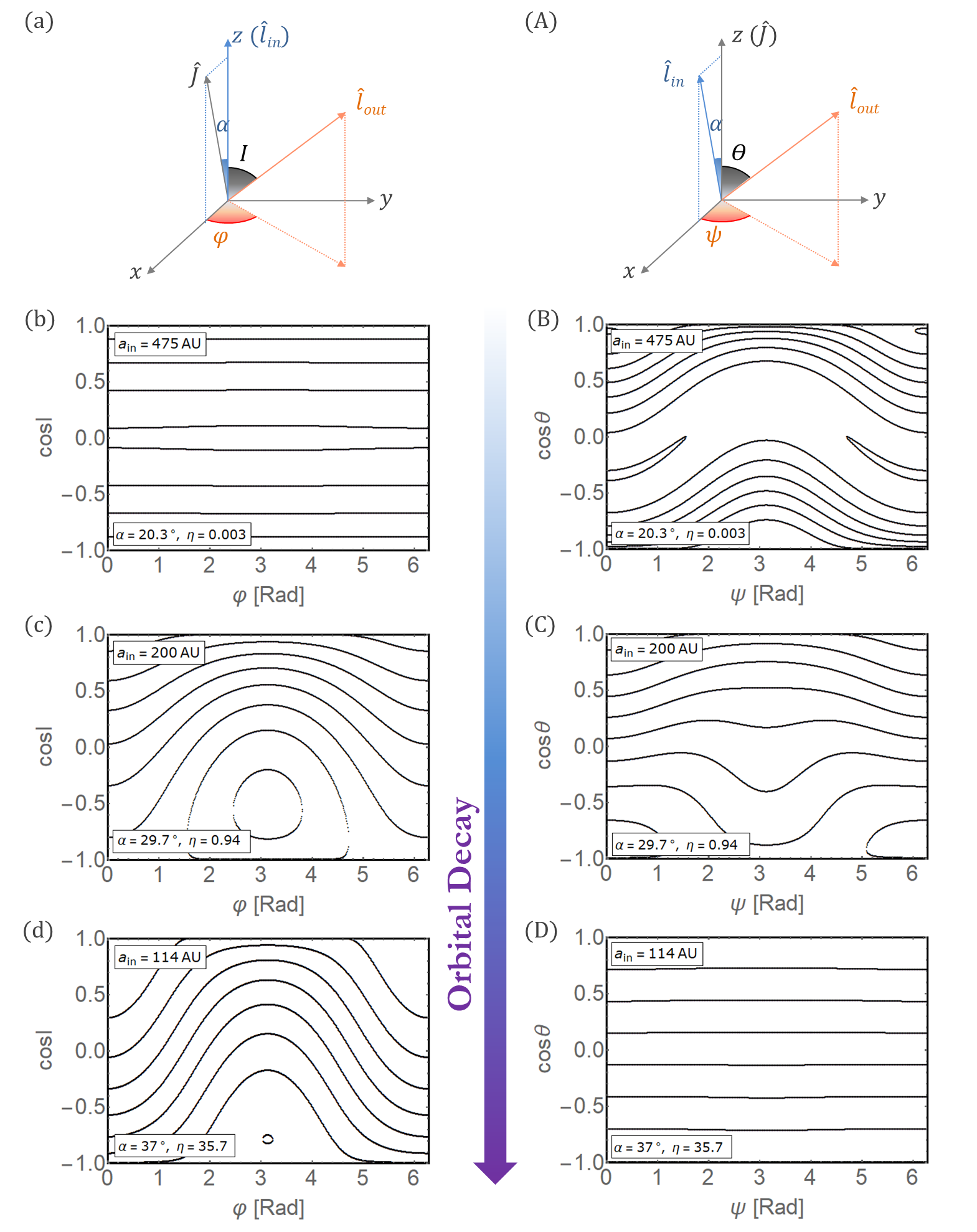}
\caption{Different types of $\hat {\bm{l}}$ behaviors
for three values of $a_\IN$ (as labeled), representing the different stages of the orbital decay of the inner BHB.
Panels (a) and (A) show the coordinate system used to describe the triple system,
where $z-$axis is aligned with $\hat {\bm{l}}_\IN$ and $\hat {\bm{J}}$, respectively.
Panels (b)-(d) and (B)-(D) show the phase-space portraits with two different sets of canonical
variables ($\cos I-\varphi$ and $\cos\theta-\psi$).
The system parameters studied here are
$m_1=4\times10^6 M_\odot$, $m_2=10^5 M_\odot$, $a_\OUT=1000\au$ and $e_\IN=e_\OUT=0$.
The solid lines shown in the panels (b)-(d) and (B)-(D)
are contours of constant $\bar{\mathcal{H}}$ (see Equations \ref{eq: Hamiltonian I}
and \ref{eq: Hamiltonian theta}),
where we keep a constant $\theta_{\mathrm{S},\IN}=90^\circ$
($\theta_{\mathrm{S},\IN}$ is the angle between $\hat {\bm{l}}_\IN$ and $\hat {\bm{S}}_1$).
}
\label{fig:phase space diagram}
\end{centering}
\end{figure*}

To study such behaviors,
we set up a coordinate system with $\hat z=\hat {\bm{l}}_\IN$,
$\hat y\sin\alpha\equiv\hat {\bm{l}}_\IN\times\hat {\bm{J}}$, and let
$\hat {\bm{l}}=\sin I(\cos\varphi\hat x+\sin\varphi\hat y)+\cos I\hat z$,
where $\alpha$ is the angle between $\hat {\bm{l}}_\IN$ and $\hat {\bm{J}}$
(see panel (a) of Figure \ref{fig:phase space diagram}).
Equation (\ref{eq: Hamiltonian dimensionless}) becomes (neglecting the $\lambda$ term)
\be\label{eq: Hamiltonian I}
\bar{\mathcal{H}}=-\frac{1}{2}\cos^2 I-\eta\Big(\cos\alpha\cos I+\sin\alpha\sin I\cos\varphi\Big).
\ee

Alternatively, we can also set up a coordinate system with $\hat z=\hat {\bm{J}}$,
as shown in the panel (A) of Figure \ref{fig:phase space diagram}.
In this case, we have
\be\label{eq: Hamiltonian theta}
\bar{\mathcal{H}}=-\frac{1}{2}\Big(\cos\alpha\cos\theta+\sin\alpha\sin\theta\cos\psi\Big)^2
-\eta\cos\theta.
\ee

Figure \ref{fig:phase space diagram} shows
the dynamical behaviors of $\hat {\bm{l}}=\hat {\bm{l}}_\OUT$
for different values of $a_\IN$, representing different stages of the orbital decay of the BHB
($m_1=4\times10^6 M_\odot$, $m_2=10^5 M_\odot$).
For each $a_\IN$, the evolution of $\hat {\bm{l}}$ follows the trajectory of
constant $\bar{\mathcal{H}}$ (using Equations \ref{eq: Hamiltonian I} or \ref{eq: Hamiltonian theta}
with a fixed spin-orbit misalignment angle).
Panels (b)-(d) and (B)-(D) show some example trajectories in the ($I,\varphi$) and ($\theta,\psi$) spaces.
We see that in the early stage (when $a_\IN=475$AU), $I$ is nearly constant
(panel b) since $\eta\ll1$; in the later stage ($a_\IN=114$AU),
$\theta$ becomes nearly constant since $\eta\gg1$ (panel D).
In between, both $I$
and $\theta$ can undergo oscillations (panels c and C).
Thus,
the outer angular momentum axis $\hat {\bm{l}}$ indeed shows three types behaviors as the
inner BHB decays.

\begin{figure*}
\begin{centering}
\includegraphics[width=13cm]{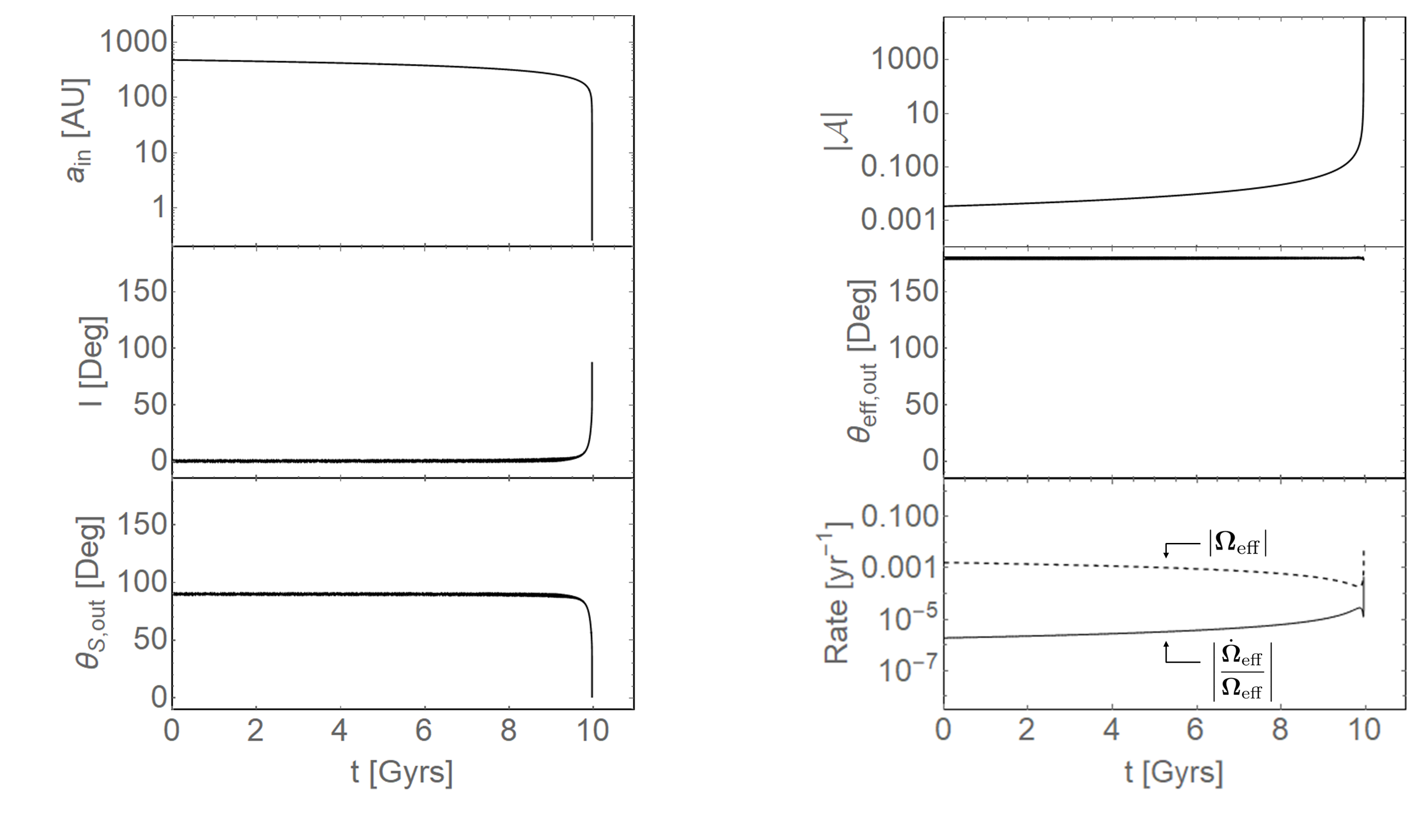}
\caption{Evolution of an outer stellar orbit around
an inner merging BHB,
where the results are obtained by integrating
Equations (\ref{eq:Full Kozai 1}), (\ref{eq:spin}), (\ref{eq:Full Kozai 3})
and (\ref{eq:Full Kozai 4}).
The system parameters are
$m_1=4\times10^6 M_\odot$, $m_2=10^5 M_\odot$, $a_\OUT=1000\au$,
$e_\IN=e_\OUT=0$ and the initial $a_\IN=475\au$, $I_0=0^\circ$.
The primary component ($m_1$) has a misaligned spin at the initial moment,
with $\theta_{\mathrm{S},\IN}=90^\circ$.
The left panels show the semimajor axis of the inner BHB ($a_\IN$), inclination $I$
(the angle between $\hat{\bm{l}}_\IN$ and $\hat{\bm{l}}$),
and misalignment $\theta_{\mathrm{S},\OUT}$ (the angle between
$\hat{\bm{S}}_1$ and $\hat{\bm{l}}$), and the right panels
show the parameter $\mathcal{A}$ (Equation \ref{eq: adiabaticity parameter}),
$\theta_{\eff,\OUT}$ (the angle between $\bm{\Omega}_\eff$ and $\hat{\bm{l}}$)
and the relevant rates for evaluating adiabaticity
(see Equation \ref{eq: adiabatic condition}).
}
\label{fig:evolution}
\end{centering}
\end{figure*}

\subsection{Final Spin-Orbit Misalignment Angles $\theta_{\mathrm{S},\OUT}^\mathrm{f}$}
\label{sec 3 2}

We now include GW dissipation of the BHB.
We expect that after the merger, $\hat{\bm{J}}\rightarrow \hat{\bm{S}}_1$
and $\theta\rightarrow \theta_{\mathrm{S}, \OUT}$.

Figure \ref{fig:evolution} shows an example of the evolution of $\hat {\bm{l}}$
during the orbital decay of BHB (with Equation \ref{eq:GW 1} included in the calculation).
In the top right panel, we introduce
\be\label{eq: adiabaticity parameter}
\mathcal{A}=\frac{\Omega_\IN}{\Omega_\OUT^{(\mathrm{N})}\cos I}=\frac{\eta}{\cos I}.
\ee
We see that the system goes through the transition
from the ``$\eta\ll1$" regime to the
``$\eta\gg1$" regime. The orientation of $\hat{\bm{l}}$ varies a lot
during the transition, and the
initial alignment of $\hat{\bm{l}}_\IN$ and $\hat{\bm{l}}$ is changed to the final alignment
of $\hat{\bm{S}}_1$ and $\hat{\bm{l}}$.

The final spin-orbit misalignment ($\theta_{\mathrm{S},\OUT}^\mathrm{f}$)
between $\hat{\bm{l}}$ and $\hat{\bm{S}}_1$
can be calculated analytically using the principle of adiabatic invariance,
if the inner binary remains circular throughout the evolution.
Equation (\ref{eq:LOUT rot}) shows that
$\bm{l}$ rotates around $\bm{\Omega}_\eff$, where
\be\label{eq: Omega eff}
\bm{\Omega}_\eff
=-\Omega_\OUT^\mathrm{(N)}(\hat {\bm{l}}_\IN\cdot \hat {\bm{l}})\hat {\bm{l}}_\IN
-\Omega_\IN\hat{\bm{J}}.
\ee
In the presence of GW dissipation, when the rate of change of $\bm{\Omega}_\eff$
is much smaller than $|\bm{\Omega}_\eff|$, i.e.,
\be\label{eq: adiabatic condition}
\bigg|\frac{\dot{\bm{\Omega}}_\eff}{\bm{\Omega}_\eff}\bigg|\ll|\bm{\Omega}_\eff|
\ee
$\bm{\Omega}_\eff$ becomes a slowly changing vector, and
the angle between $\bm{\Omega}_\eff$ and $\hat{\bm{l}}$ is expected to be
an adiabatic invariant, i.e.,
\be\label{eq: adiabatic invariant}
\theta_{\eff,\OUT}\simeq \mathrm{constant}\qquad ({\rm adiabatic~invariant}).
\ee
After the inner binary has decayed, we have
$|\Omega_\OUT^\mathrm{(N)}|\ll|\Omega_\IN|$, and $\bm{\Omega}_\eff\simeq \Omega_\IN\hat{\bm{J}}$.
Therefore,
\be\label{eq: General final angle}
\theta_{\mathrm{S},\OUT}^\mathrm{f}\simeq\theta_{\eff,\OUT}^\mathrm{f}
=\theta_{\eff,\OUT}^0.
\ee

To obtain $\theta_{\eff,\OUT}^0$, we note that the orientation of the initial $\bm{\Omega}_\eff$
is determined by both
$\hat{\bm{l}}_\IN$ and $\hat{\bm{J}}$.
For the outer orbits with $|\Omega_\OUT^\mathrm{(N)}|\gg|\Omega_\IN|$
(generally corresponding to the systems with small $a_\OUT$), we have
$\bm{\Omega}_\eff\simeq-\Omega_\OUT^\mathrm{(N)}\hat {\bm{l}}_\IN$.
As a result, the final spin-orbit misalignment angle is equal to
the initial inclination angle between
$\hat{\bm {l}}_\IN$ and $\bm{l}$, i.e.,
$\theta_{\mathrm{S},\OUT}^\mathrm{f}\simeq\theta_{\eff,\OUT}^0\simeq I_0$.
For the example shown in Figure \ref{fig:evolution},
we see that the adiabatic criterion (Equation \ref{eq: adiabatic condition}) is satisfied
and the adiabatic invariant $\theta_{\eff,\OUT}$ is almost a constant.
Since $\hat {\bm{l}}$ and $\hat {\bm{l}}_\IN$ are initially aligned, $I_0=0$,
the final spin-orbit misalignment angle $\theta_{\mathrm{S},\OUT}^\mathrm{f}=0$.

For the distant outer orbits, we have $|\Omega_\OUT^\mathrm{(N)}|\ll|\Omega_\IN|$, and
$\bm{\Omega}_\eff\simeq-\Omega_\IN\hat {\bm{J}}$. Therefore, we expect that
$\theta_{\mathrm{S},\OUT}^\mathrm{f}\simeq\theta_{\eff,\OUT}^0\simeq \theta_0$,
where $\theta_0$ is the angle between $\hat {\bm{J}}$ and $\hat {\bm{l}}$
at the initial moment.
For the specific configuration with $I_0=0$,
we have $\theta_{\mathrm{S},\OUT}^\mathrm{f}\simeq\alpha_0$,
where $\alpha_0$ is the initial angle between $\hat {\bm{J}}$ and $\hat {\bm{l}}_\IN$.

\section{Results for Initially Coplanar Outer Orbits}
\label{sec 4}
\subsection{Fiducial Case: $m_2=10^5M_\odot$}
\label{sec 4 1}

\begin{figure}
\begin{centering}
\includegraphics[width=8.2cm]{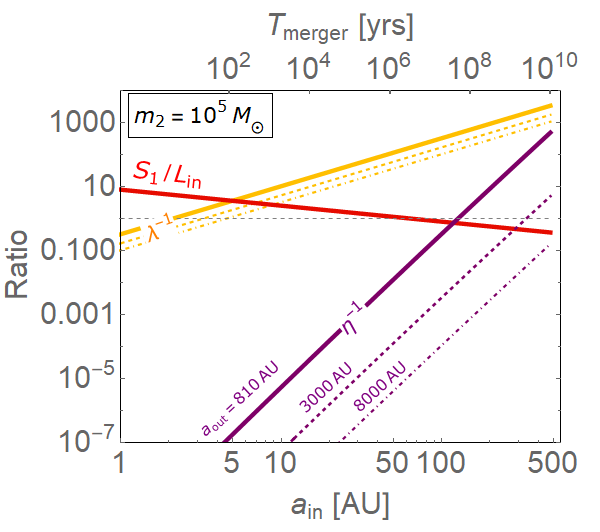}
\caption{The variation of the values of $\eta$ and $\lambda$ as the inner BHB decays.
The inner BHB has masses $m_1=4\times10^6 M_\odot$ and $m_2=10^5 M_\odot$.
The results (purple and orange lines) are obtained by using
Equation (\ref{eq: eta}) with $\theta_{\mathrm{S},\IN}=0^\circ$,
where the solid, dashed and dot-dashed lines
are for the given $a_\OUT$ (as labeled; the minimum $a_\OUT$ is evaluated by Equation \ref{eq: aoutc}). We also show the ratio of $S_1/L_\IN$
as a function of $a_\IN$.
}
\label{fig:parameter space 105}
\end{centering}
\end{figure}

We now study the evolution of the outer orbits with different radius ($a_\OUT$)
as the inner BHB decays.
We consider the initially coplanar case with $I_0=0^\circ$ and $m_2=10^5M_\odot$ in this section.

Figure \ref{fig:parameter space 105} shows $\lambda^{-1}$ and $\eta^{-1}$ (see Equation \ref{eq: eta})
as a function of $a_\IN$ for a given $a_\OUT$.
The values of $\eta$ are obtained by setting $\theta_{\mathrm{S},\IN}=0^\circ$.
We find that the nodal precession induced by GR ($\Omega_\OUT^{'(\gr)}$)
is always weaker than the
Newtonian one ($\Omega_\OUT^{(\mathrm{N})}$), until the inner BH binary
has become sufficiently compact.
On the other hand, when the BHB is wide, the systems, especially
for the close test particle orbits (e.g., $a_\OUT=810\au$), are
in the ``$\eta\gg1$" regime, in which the
Newtonian precession of $\hat {\bm{l}}$ around $\hat {\bm{l}}_\IN$
is much stronger than the
precession of $\hat {\bm{l}}_\IN$ around $\hat {\bm{J}}$.
This implies that the direction of $\bm{\Omega}_\eff$
is approximately parallel to $\hat {\bm{l}}_\IN$
and $\theta_{\eff,\OUT}^0\simeq I_0$.
However, if the test particle is further away from the central BHB
(i.e., $a_\OUT>3000\au$),
$\eta$ is close to unity and
the orientation of $\bm{\Omega}_\eff$
is determined by both $\hat {\bm{l}}_\IN$ and $\hat {\bm{J}}$.

\begin{figure*}
\begin{centering}
\includegraphics[width=16cm]{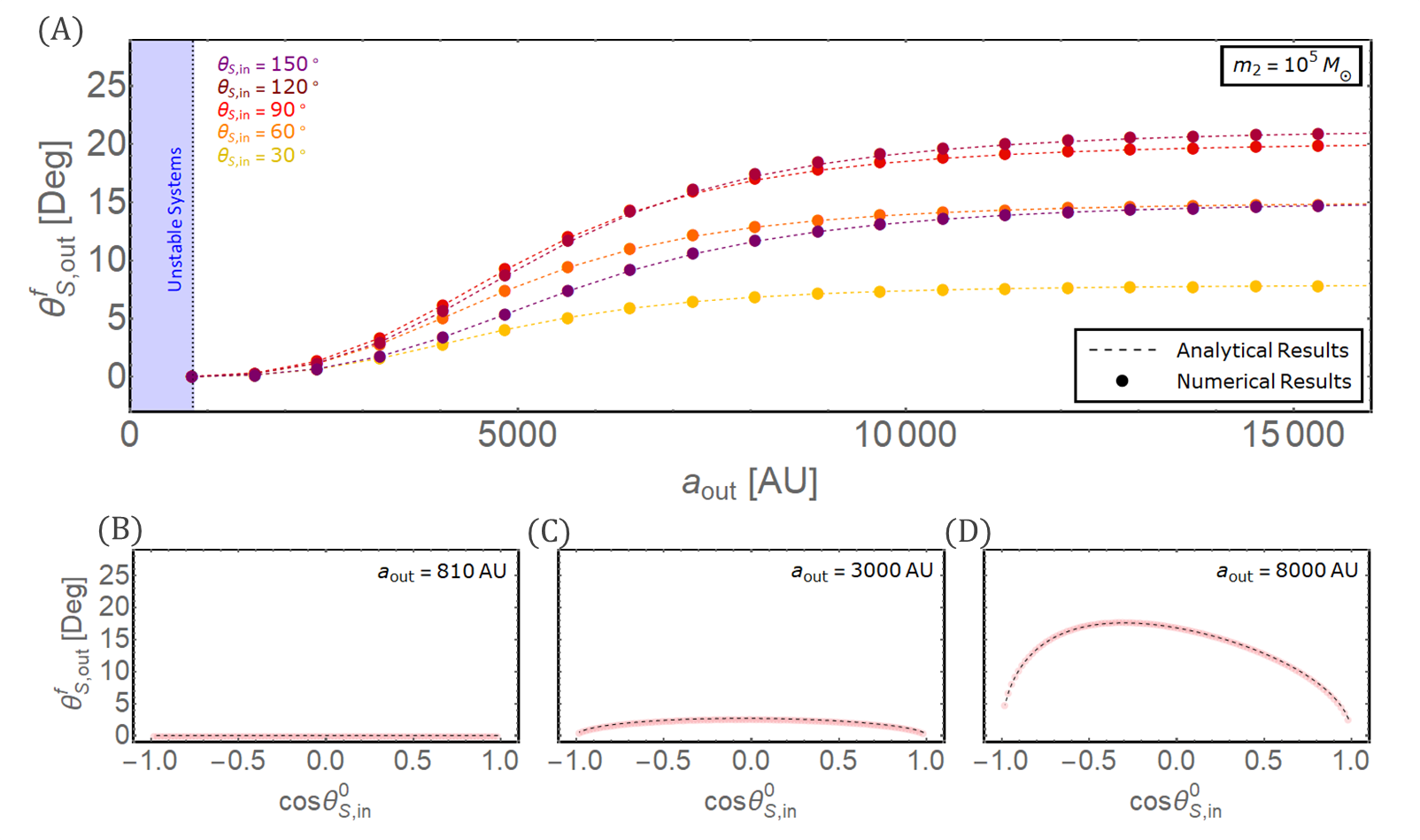}
\caption{Panel (A) shows the final spin-orbit misalignment angles
$\theta_{\mathrm{S},\OUT}^\mathrm{f}$ as a function of $a_\OUT$,
for different initial spin orientations (as labeled).
The system parameters follow the example shown in Figure \ref{fig:parameter space 105}.
The stability criterion is given by Equation (\ref{eq: aoutc}).
All the dots are the numerical results obtained by integrating Equations (\ref{eq:Full Kozai 1}),
(\ref{eq:spin}), (\ref{eq:Full Kozai 3}) and (\ref{eq:Full Kozai 4}).
The dashed lines are the analytical results
based on Equation (\ref{eq: Omega eff}).
Panels (C)-(D) show the final angles $\theta_{\mathrm{S},\OUT}^\mathrm{f}$
as a function of a full range of $\cos\theta_{\mathrm{S},\IN}^0$,
with three values of $a_\OUT$. Again, the dots and the dashed lines
are obtained numerically and analytically, respectively.
}
\label{fig:aout dependence 105}
\end{centering}
\end{figure*}

In Figure \ref{fig:aout dependence 105}, panel (A) shows
the final spin-orbit angles $\theta_{\mathrm{S},\OUT}^\mathrm{f}$
for a series of test particle orbits with different separations,
for several values of $\theta_{\mathrm{S},\IN}^0$.
We obtain the numerical results (dots) by integrating
Equations (\ref{eq:Full Kozai 1}), (\ref{eq:spin}), (\ref{eq:Full Kozai 3})
and (\ref{eq:Full Kozai 4})
and the analytical results based on Equation (\ref{eq: General final angle}).
We find that the analytic prediction (dashed lines) agrees well
with the numerical results.
For the close test particle orbits,
the final angular momentum $\hat {\bm{l}}$ always points in the direction of
the spin $\hat {\bm{S}}_1$, i.e., $\theta_{\mathrm{S},\OUT}^\mathrm{f}\simeq I_0=0^\circ$,
regardless of the initial spin orientation.
This is because $\bm{\Omega}_\eff\varpropto\hat {\bm{l}}_\IN$
for the orbits with $a_\OUT\lesssim3000\au$
(as shown in Figure \ref{fig:parameter space 105}).
On the other hand, for $a_\OUT\gtrsim3000\au$,
the final angle $\theta_{\mathrm{S},\OUT}^\mathrm{f}$
is only determined by $\theta_{\eff,\OUT}^0$,
and $\theta_{\mathrm{S},\OUT}^\mathrm{f}\simeq\alpha_0$ (the angle between $\hat {\bm{J}}$ and $\hat {\bm{l}}_\IN$)
as $a_\OUT\gtrsim10^4\au$.
Since the initial orientation of $\bm{J}$ depends on $\theta_{\mathrm{S},\IN}^0$,
we see that the angles $\theta_{\mathrm{S},\OUT}^\mathrm{f}$ corresponding to different
$\theta_{\mathrm{S},\OUT}^\mathrm{f}$ differ at large $a_\OUT$.

\begin{figure*}
\begin{centering}
\includegraphics[width=8.2cm]{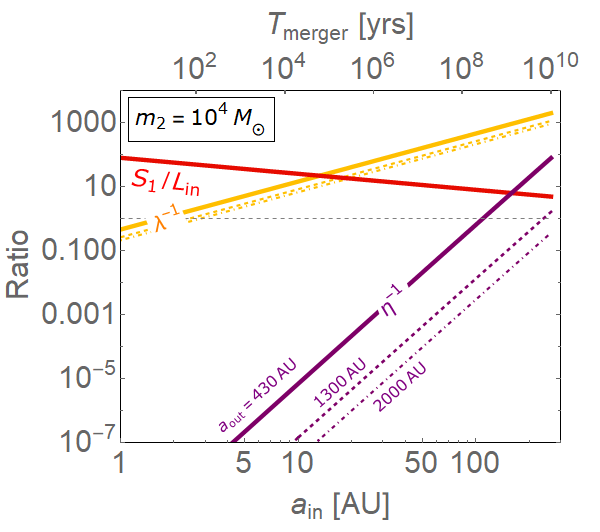}
\includegraphics[width=8.2cm]{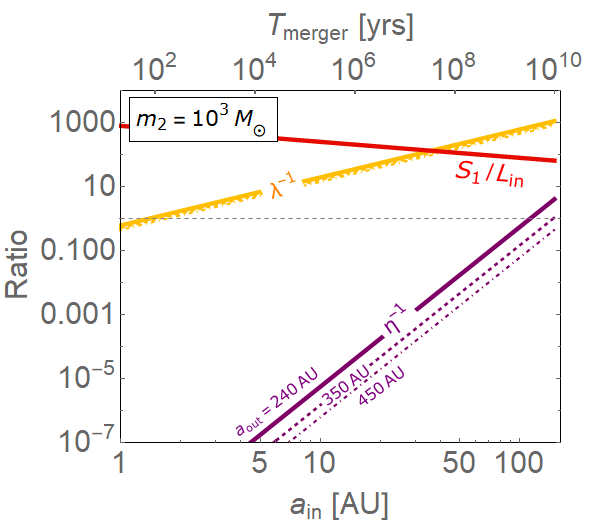}
\caption{Same as Figure \ref{fig:parameter space 105},
except for $m_2=10^4M_\odot$ (left panel) and $m_2=10^3M_\odot$ (right panel).
}
\label{fig:parameter space 1043}
\end{centering}
\end{figure*}

\begin{figure*}
\begin{centering}
\includegraphics[width=16cm]{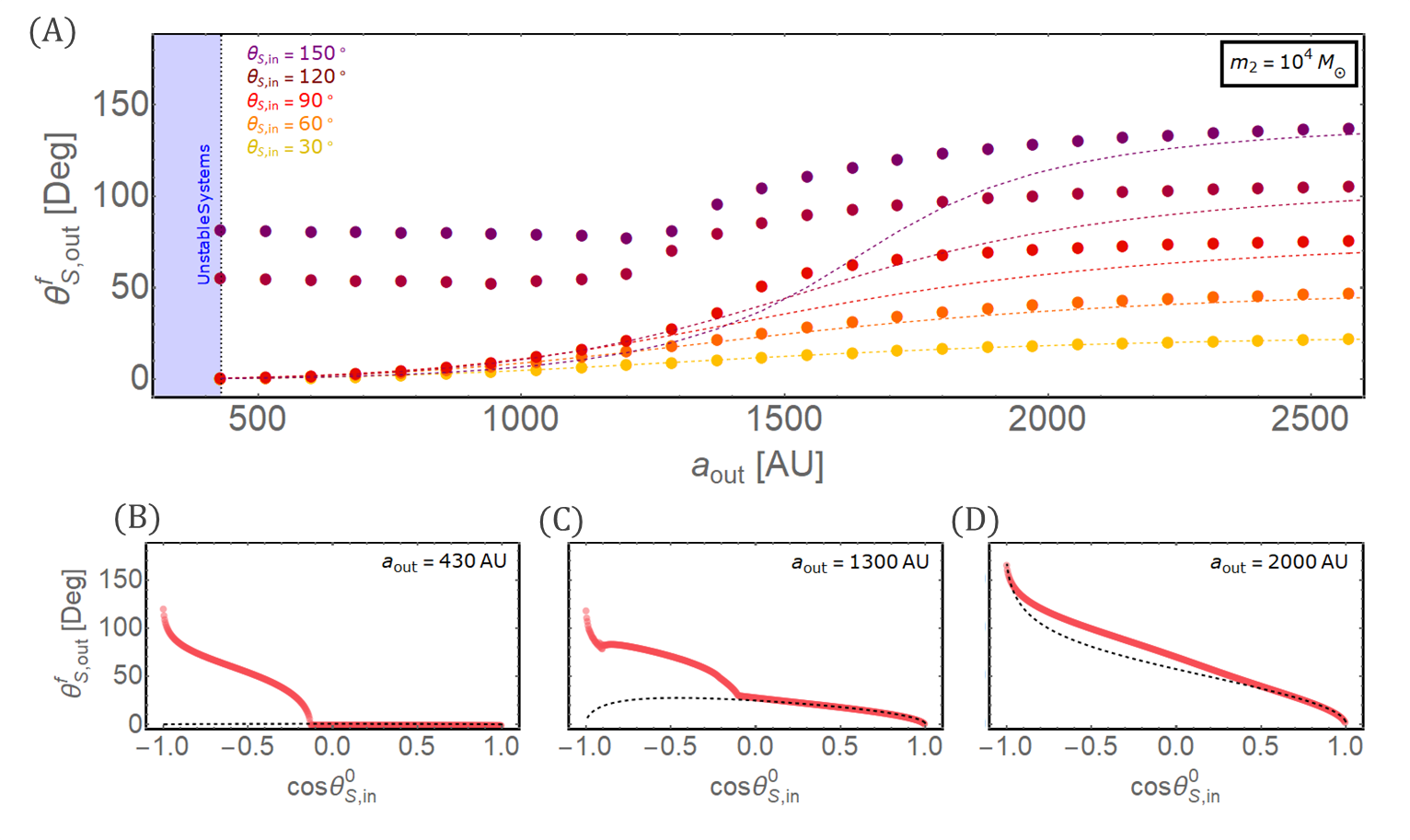}
\caption{Same as Figure \ref{fig:aout dependence 105},
except for $m_2=10^4M_\odot$.
}
\label{fig:aout dependence 104}
\end{centering}
\end{figure*}

\begin{figure*}
\begin{centering}
\includegraphics[width=8.5cm]{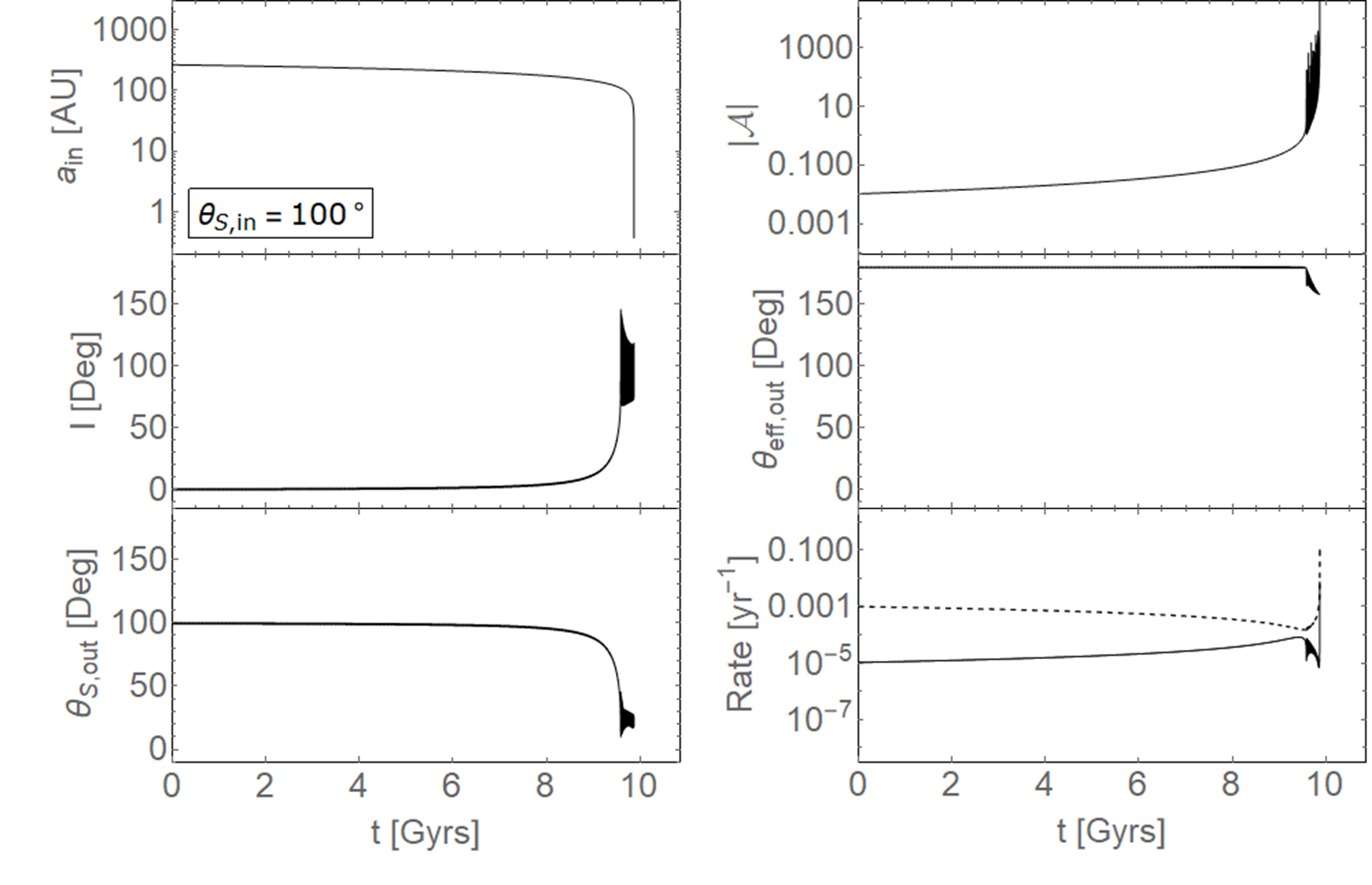}~~~~~~~~~~~
\includegraphics[width=8.5cm]{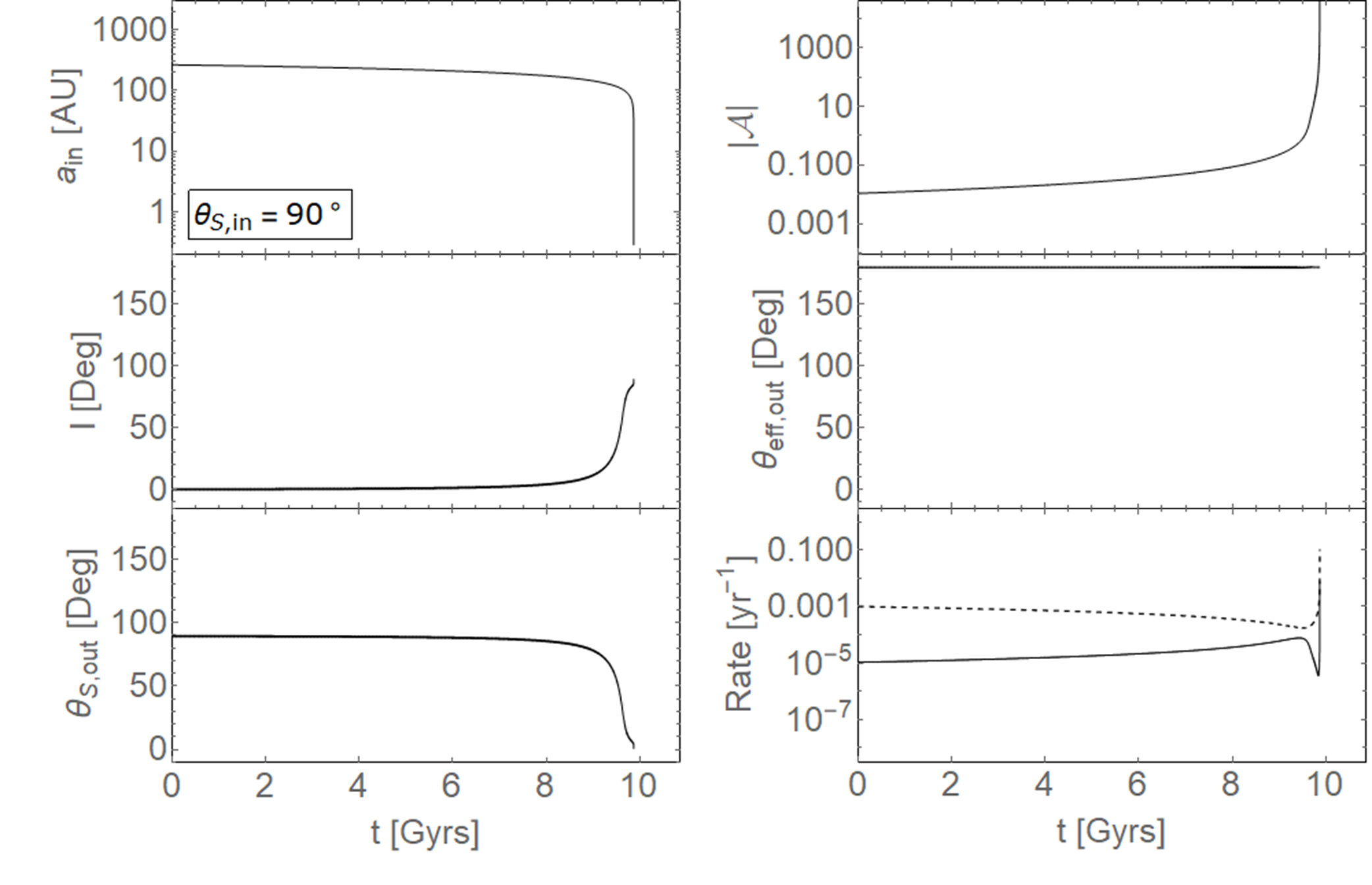}
\caption{Similar to Figure \ref{fig:evolution}, but
the system parameters here are
$m_1=4\times10^6 M_\odot$, $m_2=10^4 M_\odot$,
$a_\IN=265\au$, $a_\OUT=430\au$, $e_\IN=e_\OUT=0$ and $I_0=0^\circ$.
We consider two values of $\theta_{\mathrm{S},\IN}$
(as labeled) in the left and right panels.
}
\label{fig:evolution 104}
\end{centering}
\end{figure*}

Panels (B)-(D) of Figure \ref{fig:aout dependence 105} show the dependence of
$\theta_{\mathrm{S},\OUT}^\mathrm{f}$
on $\theta_{\mathrm{S},\IN}^0$ for three values of $a_\OUT$.
We find that
the analytical results are in excellent agreement
with the numerical calculations.

\begin{figure*}
\begin{centering}
\includegraphics[width=16cm]{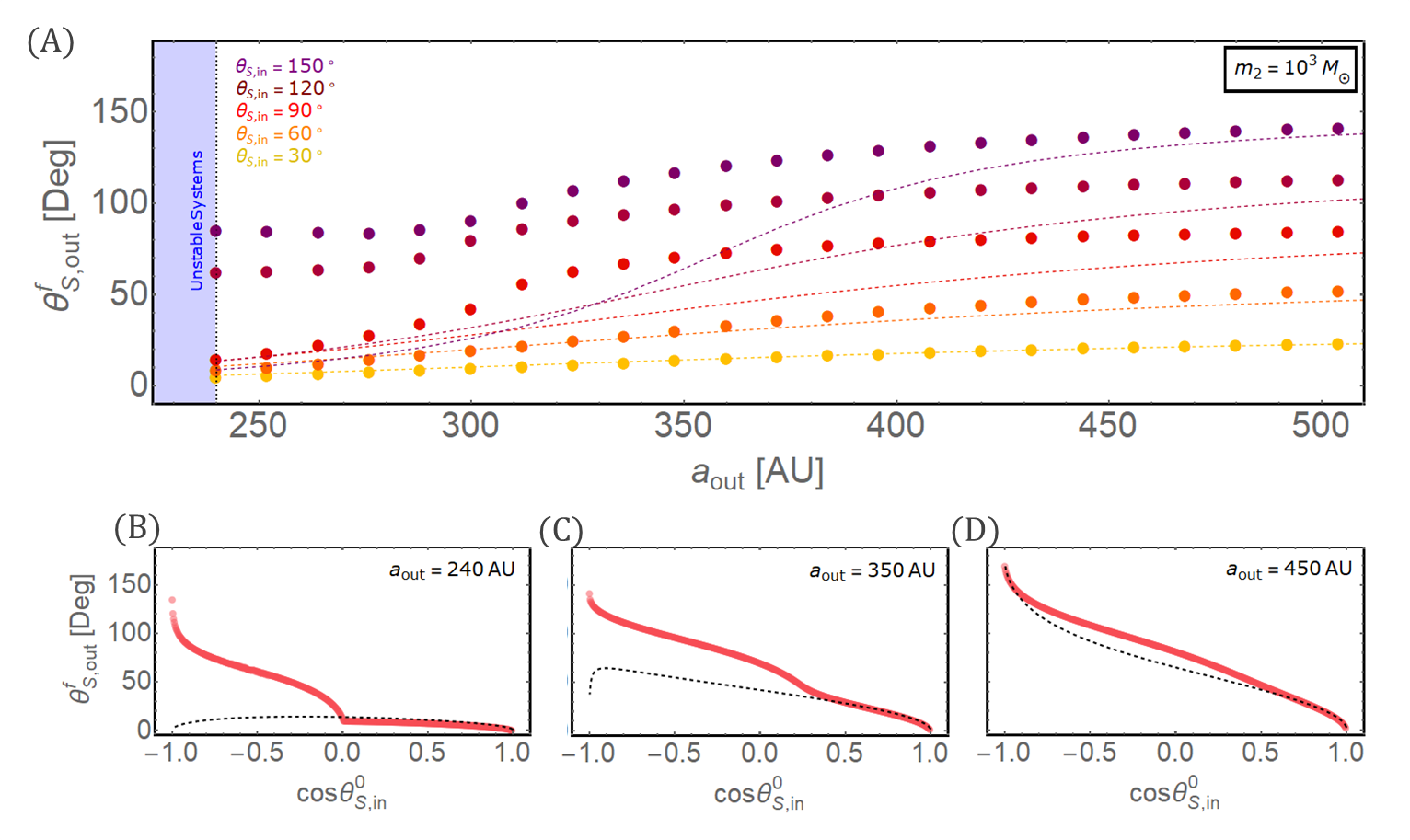}
\caption{Same as Figure \ref{fig:aout dependence 105},
except for $m_2=10^3M_\odot$.
}
\label{fig:aout dependence 103}
\end{centering}
\end{figure*}

\begin{figure*}
\begin{centering}
\includegraphics[width=8.5cm]{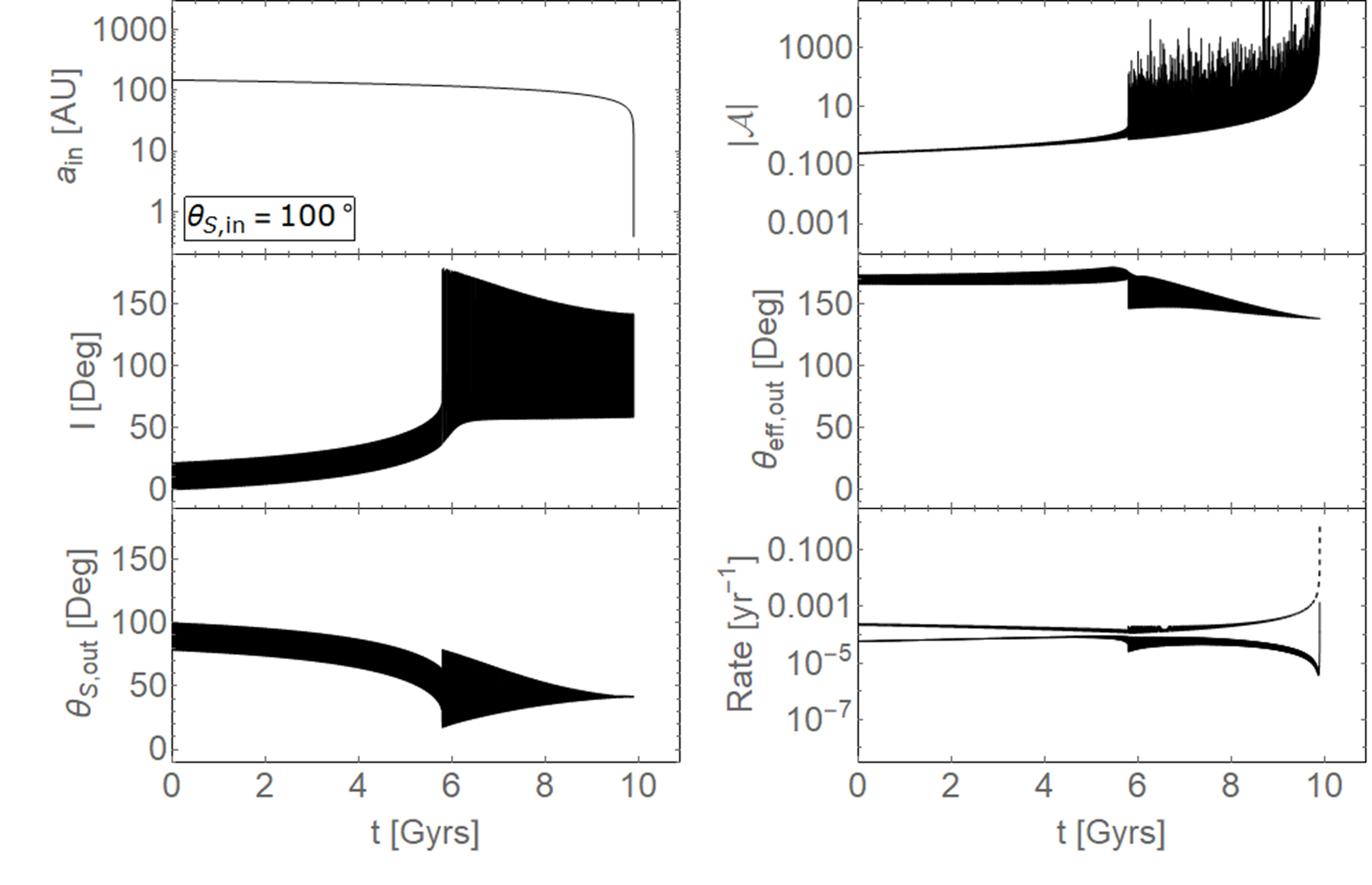}~~~~~~~~~~~
\includegraphics[width=8.5cm]{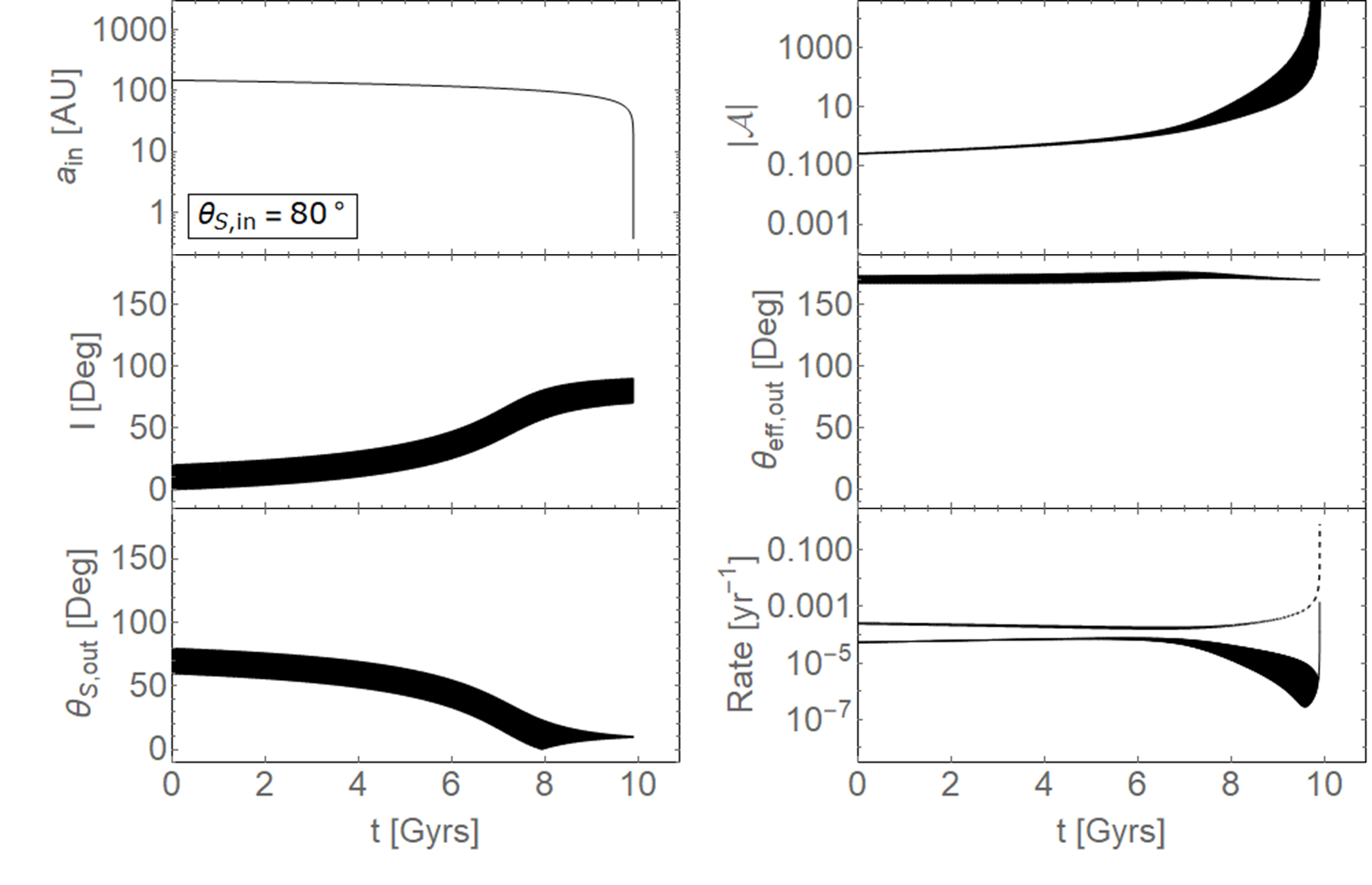}
\caption{Similar to Figure \ref{fig:evolution}, but
the system parameters here are
$m_1=4\times10^6 M_\odot$, $m_2=10^3 M_\odot$, $a_\IN=149\au$,
$a_\OUT=240\au$, $e_\IN=e_\OUT=0$ and $I_0=0^\circ$.
We consider two values of $\theta_{\mathrm{S},\IN}$
(as labeled) in the left and right panels.
}
\label{fig:evolution 103}
\end{centering}
\end{figure*}

\subsection{$m_2=10^4M_\odot$ and $m_2=10^3M_\odot$}
\label{sec 4 2}

If $m_2$ becomes lighter, in order to have BHB merging
within the Hubble timescale, $a_\IN$ should be smaller
(as shown in Figure \ref{fig:parameter space 105}).
The initial systems maybe close to or even already
in the ``$\eta\sim1$" regime,
indicating that the angular momentum of the close
test particle orbit $\hat {\bm{l}}$ may experience more complicated evolution
at the early stage of the merger of the inner BHB.

\begin{figure*}
\begin{centering}
\includegraphics[width=16cm]{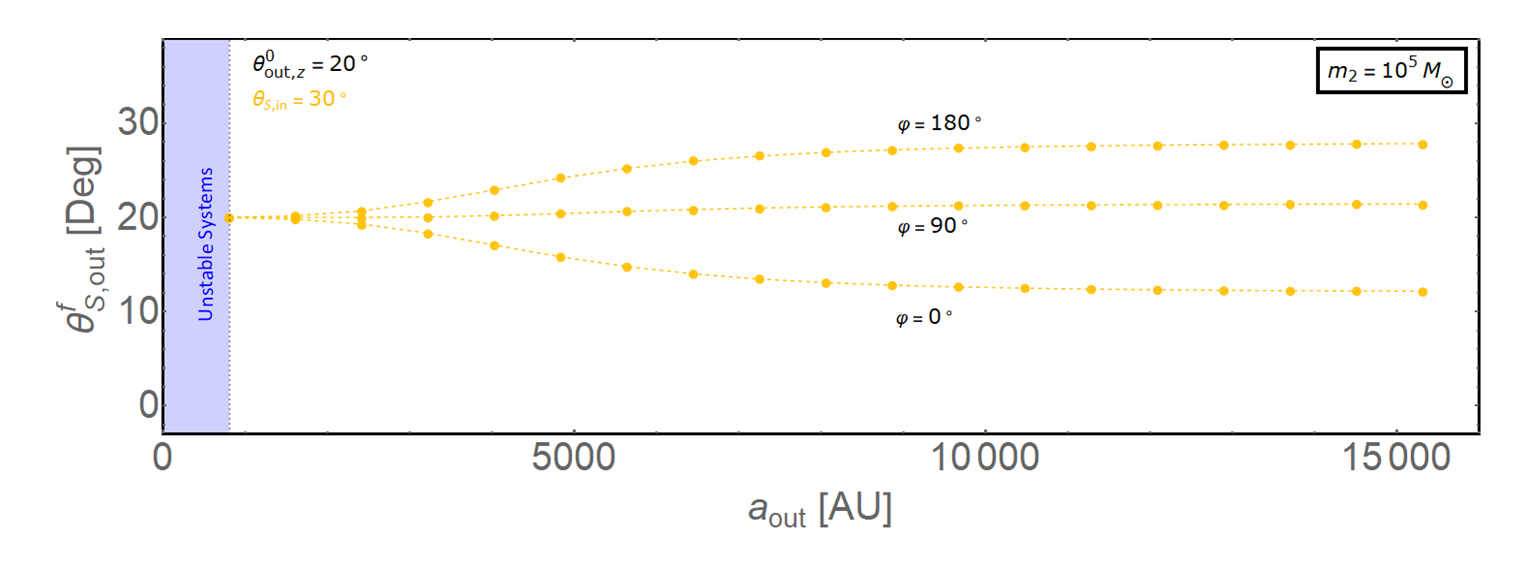}
\caption{The final spin-orbit misalignment angle
$\theta_{\mathrm{S},\OUT}^\mathrm{f}$ as a function of $a_\OUT$.
The system parameters are $m_1=4\times10^6 M_\odot$, $m_2=10^5 M_\odot$,
$a_\IN=475\au$, and $e_\IN=e_\OUT=0$.
The primary component of BHB has a misaligned spin direction,
with $\theta_{\mathrm{S},\IN}^0=30^\circ$.
The angular momentum of the outer orbit is initialized inclined with respect to
the initial direction of $\hat {\bm{l}}_\IN$
($z-$axis; see also the panel (a) of Figure \ref{fig:phase space diagram})
by $20^\circ$ (i.e., $\theta_{\OUT, \mathrm{z}}^0=20^\circ$), with three different phase angles
($\varphi=0^\circ, 90^\circ, 180^\circ$).
The dots are from the numerical calculation and the dashed lines
are the analytical results (i.e., $\theta_{\mathrm{S},\OUT}^\mathrm{f}=\theta_{\eff,\OUT}^0$).
}
\label{fig:aout dependence inclined Lout 30 Phase}
\end{centering}
\end{figure*}

Figure \ref{fig:parameter space 1043} shows how $\lambda$
and $\eta$ change as $a_\IN$ decreases
when $m_2=10^4M_\odot$ (left panel) and $m_2=10^3M_\odot$ (right panel).
Here, since $S_1\gg L_\IN$,
the orientation of $\hat {\bm{J}}$ is dominated by $\hat {\bm{S}}_1$.

Figure \ref{fig:aout dependence 104} shows the final angle
$\theta_{\mathrm{S},\OUT}^\mathrm{f}$ as a function of $a_\OUT$ for
a range of $\theta_{\mathrm{S},\IN}^0$ values.
Compared to the results shown in Figure \ref{fig:aout dependence 105},
the analytical predictions are only valid for the small $\theta_{\mathrm{S},\IN}^0$ or the
the distant outer orbits (see also the panel D);
for the test particle orbit with small $a_\OUT$, the analytical results
break down when $\theta_{\mathrm{S},\IN}^0\gtrsim90^\circ$ (see also panels B and C).

Figure \ref{fig:evolution 104} shows two evolution examples for a system with small $a_\OUT$.
We identify two main reasons for the discrepancy between the analytical and numerical results for $\theta_{\mathrm{S},\OUT}^\mathrm{f}$:
(i) The time of entry into ``$\eta\sim1$" regime.
The systems with small $m_2$ tend to have a relatively large $\eta$ ($\lesssim1$),
thus will enter the ``$\eta\sim1$" regime earlier.
The inclination angle $I$ shown in Figure \ref{fig:evolution 104}
has a chance to be excited (left panel) or experience oscillations (right panel)
at earlier times
compared to the example shown in Figure \ref{fig:evolution}.
Note that the exact value of $\eta$ depends on
the choice of $\theta_{\mathrm{S},\IN}^0$ (see Figure \ref{fig:parameter space});
(ii) Crossing $90^\circ$ in $I$.
For the BHB with small mass ratio, the direction of $\hat {\bm{J}}$
is dominated by the spin vector $\hat {\bm{S}}_1$
instead of $\hat {\bm{l}}_\IN$
(see Figure \ref{fig:parameter space 1043}).
Thus, for a given $\theta_{\mathrm{S},\IN}^0$,
the angle between $\hat {\bm{l}}_\IN$ and $\hat {\bm{J}}$ (i.e., $\alpha$)
is larger than the one for a BHB with comparable masses
(e.g., Figure \ref{fig:parameter space 105}).
The large $\alpha$ value may easily induce large inclinations
($I\gtrsim90^\circ$) due to the precession of $\hat {\bm{l}}_\IN$
around $\hat {\bm{J}}$ as the system reach the ``$\eta\sim1$" regime.
Therefore,
the crossing through $90^\circ$ in $I$ may occur and induces significant oscillations
in $|\mathcal{A}|$ and $|\dot{\bm{\Omega}}_\eff|$,
breaking the adiabaticity condition.

Figure \ref{fig:aout dependence 103} shows the results for $m_2=10^3M_\odot$.
Similar to Figure \ref{fig:aout dependence 104}, we find that
the analytical results are in an agreement
with the numerical calculations except when $a_\OUT$ is small
($a_\OUT\lesssim400\au$) and $\theta_{\mathrm{S},\IN}^0$
is large ($\theta_{\mathrm{S},\IN}^0\gtrsim90^\circ$).

Different from the case of $m_2=10^4M_\odot$, the system with $m_2=10^3M_\odot$
has $\eta\simeq1$ at the initial time, which means it will pass through the
``$\eta\sim1$" regime much earlier. We see in Figure \ref{fig:evolution 103} that
the inclination angle $I$ undergoes small amplitude oscillations
in the early stage, which is a result of the precession of
$\hat {\bm{l}}_\IN$ around $\hat {\bm{J}}$.
After the excitation, $I$ keeps oscillating for a long time
until the inner BHB merges.

\section{Numerical Results for Misaligned and Eccentric Outer Orbits}
\label{sec 5}
\subsection{Initially Inclined $\hat{\bm{l}}$}
\label{sec 5_1}

\begin{figure*}
\begin{centering}
\includegraphics[width=16cm]{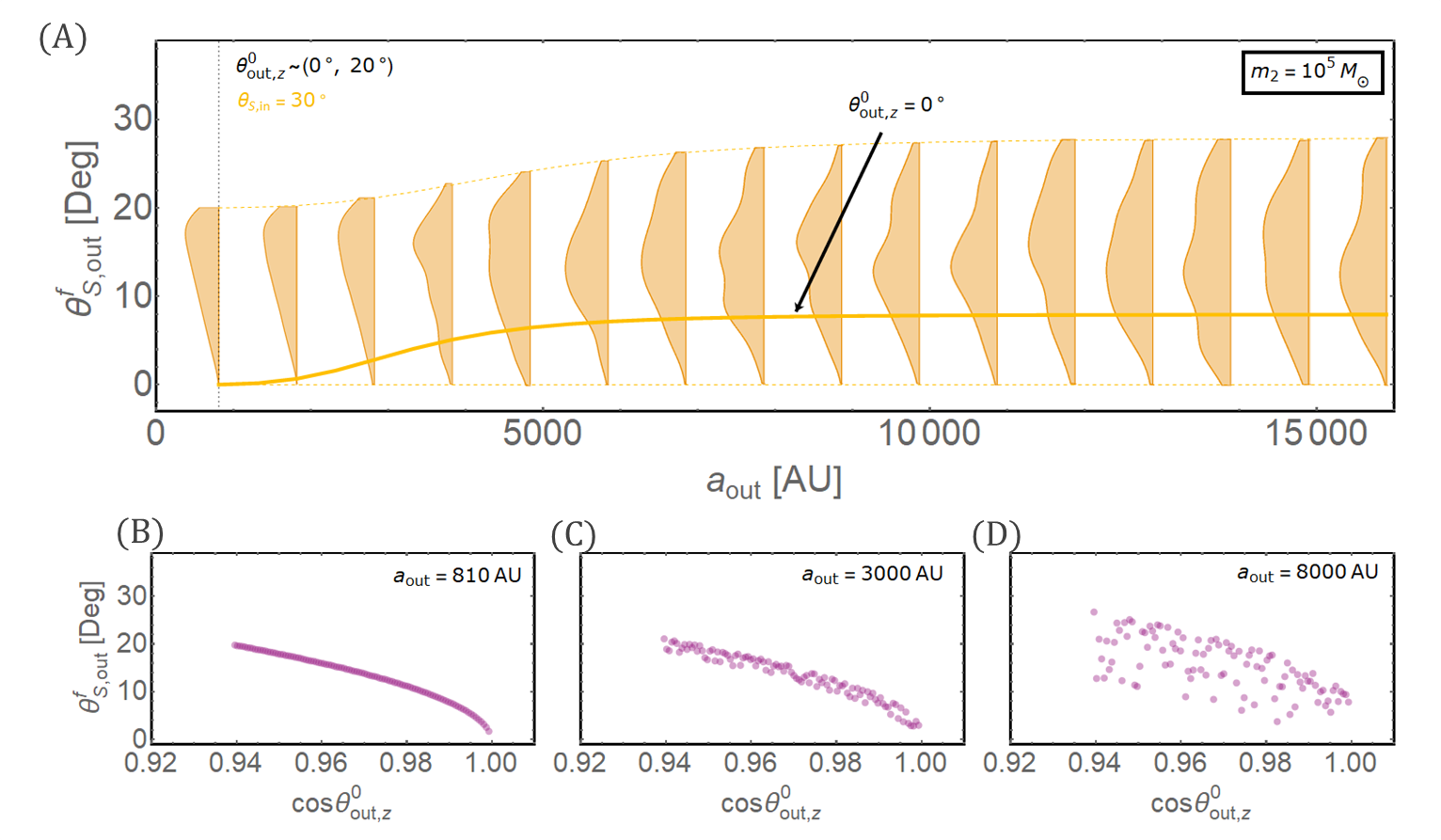}
\caption{Panel (A) shows the PDF distribution of the angles
$\theta_{\mathrm{S},\OUT}^\mathrm{f}$ for different $a_\OUT$.
The parameters are the same as in
Figure \ref{fig:aout dependence inclined Lout 30 Phase}, except
for each $a_\OUT$, we choose 100 values of $\theta_{\OUT, \mathrm{z}}$
within the range of [$0^\circ$, $20^\circ$]
(uniform in $\cos\theta_{\OUT, \mathrm{z}}$)
and the initial phase angle is set to be randomly distributed
from $0$ to $2\pi$. The dashed lines are given by $\theta_{\mathrm{S},\OUT}^\mathrm{f}=0^\circ$ and
$\theta_{\mathrm{S},\OUT}^\mathrm{f}=\theta_{\eff,\OUT}^0$ with the initial
$\theta_{\OUT, \mathrm{z}}^0=20^\circ$ at $\varphi=\pi$.
We highlight the results
from $\theta_{\OUT, \mathrm{z}}^0=0^\circ$ as the solid line.
Panels (B)-(D) show the distribution of $\theta_{\mathrm{S},\OUT}^\mathrm{f}$
as a function of $\cos\theta_{\OUT, \mathrm{z}}$,
for three values of $a_\OUT$.
}
\label{fig:aout dependence inclined Lout 30}
\end{centering}
\end{figure*}

We now consider the general case in which
$\hat {\bm{l}}$ is not aligned with $\hat {\bm{l}}_\IN$ initially,
focusing on systems with $m_1=4\times10^6 M_\odot$, $m_2=10^5 M_\odot$.

Figure \ref{fig:aout dependence inclined Lout 30 Phase}
shows our results when the initial $\hat {\bm{l}}$ is inclined to
$\hat {\bm{l}}_\IN$ by $20^\circ$.
We find that the analytical results for $\theta_{\mathrm{S},\OUT}^\mathrm{f}$
agree well with
the numerical results. In addition, we see that three lines
from different initial phase angles converge into a single line at small $a_\OUT$.
This is because in this case,
$\bm{\Omega}_\eff\simeq-\Omega_\OUT^\mathrm{(N)}(\hat {\bm{l}}_\IN\cdot\hat {\bm{l}})\hat {\bm{l}}_\IN$,
and $\theta_{\mathrm{S},\OUT}^\mathrm{f}$ is only determined by $I_0$
instead of $\varphi$.
If $a_\OUT$ is sufficient large,
$\bm{\Omega}_\eff\simeq-\Omega_\IN\hat {\bm{J}}$
and $\theta_{\mathrm{S},\OUT}^\mathrm{f}=\theta_0$,
which depends on the initial phase angle.
As seem in the panel (A) of Figure \ref{fig:phase space diagram},
the minimum and maximum values of
$\theta_0$ can be achieved when $\varphi=0^\circ, 180^\circ$, respectively.
Therefore, the range of $\theta_{\mathrm{S},\OUT}^\mathrm{f}$
can be well characterized for the distant test-particle orbits.

\begin{figure*}
\begin{centering}
\includegraphics[width=16cm]{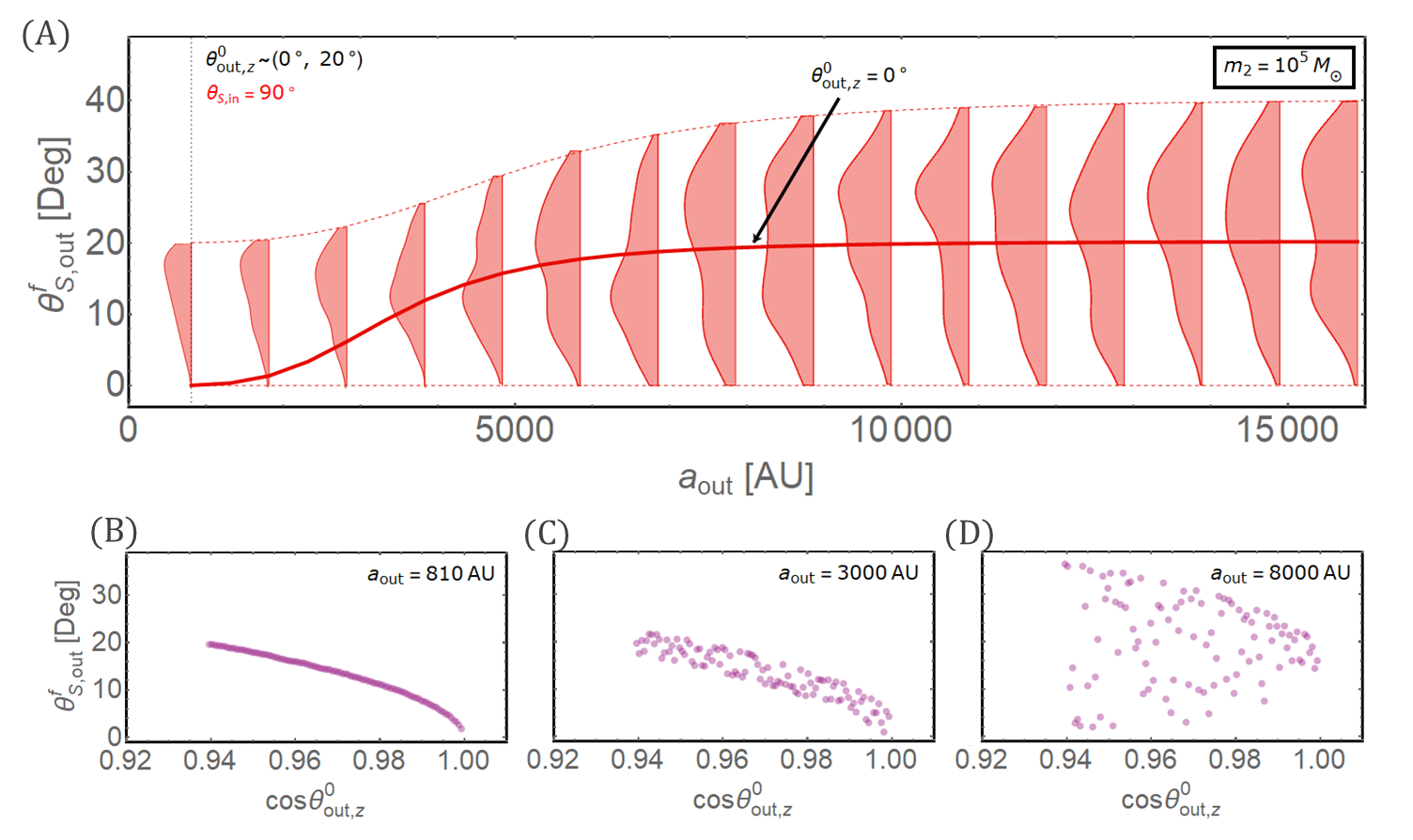}
\caption{Same as Figure \ref{fig:aout dependence inclined Lout 30},
except for $\theta_{\mathrm{S},\IN}^0=90^\circ$.
}
\label{fig:aout dependence inclined Lout 90}
\end{centering}
\end{figure*}

To determine the final orientation of a stellar disk
with finite ``thickness",
we consider a range of initially inclined $\hat {\bm{l}}$
with misalignment angle
$\theta_{\OUT, \mathrm{z}}\in(0^\circ, 20^\circ)$
($\theta_{\OUT, \mathrm{z}}$ is the angle between $\hat {\bm{l}}$ and $z-$axis, i.e., initial $\hat {\bm{l}}_\IN$) at each $a_\OUT$.
For each $I_0$, we consider a random phase $\varphi$ from $0$ to $2\pi$.
The results are shown in Figure \ref{fig:aout dependence inclined Lout 30}.
A wide range of $\theta_{\mathrm{S},\OUT}^\mathrm{f}$ are produced for a given $a_\OUT$.

To characterize the role of the initial spin orientation,
we perform the similar calculations
with $\theta_{\mathrm{S},\IN}^0=90^\circ$.
The results are shown in Figure \ref{fig:aout dependence inclined Lout 90}.
Compared to Figure \ref{fig:aout dependence inclined Lout 30},
the distribution of $\theta_{\mathrm{S},\OUT}^\mathrm{f}$ is widened,
but all $\theta_{\mathrm{S},\OUT}^\mathrm{f}$ values
are within $40^\circ$.

\subsection{Eccentric Outer Orbits}
\label{sec 5_2}

\begin{figure*}
\begin{centering}
\includegraphics[width=16cm]{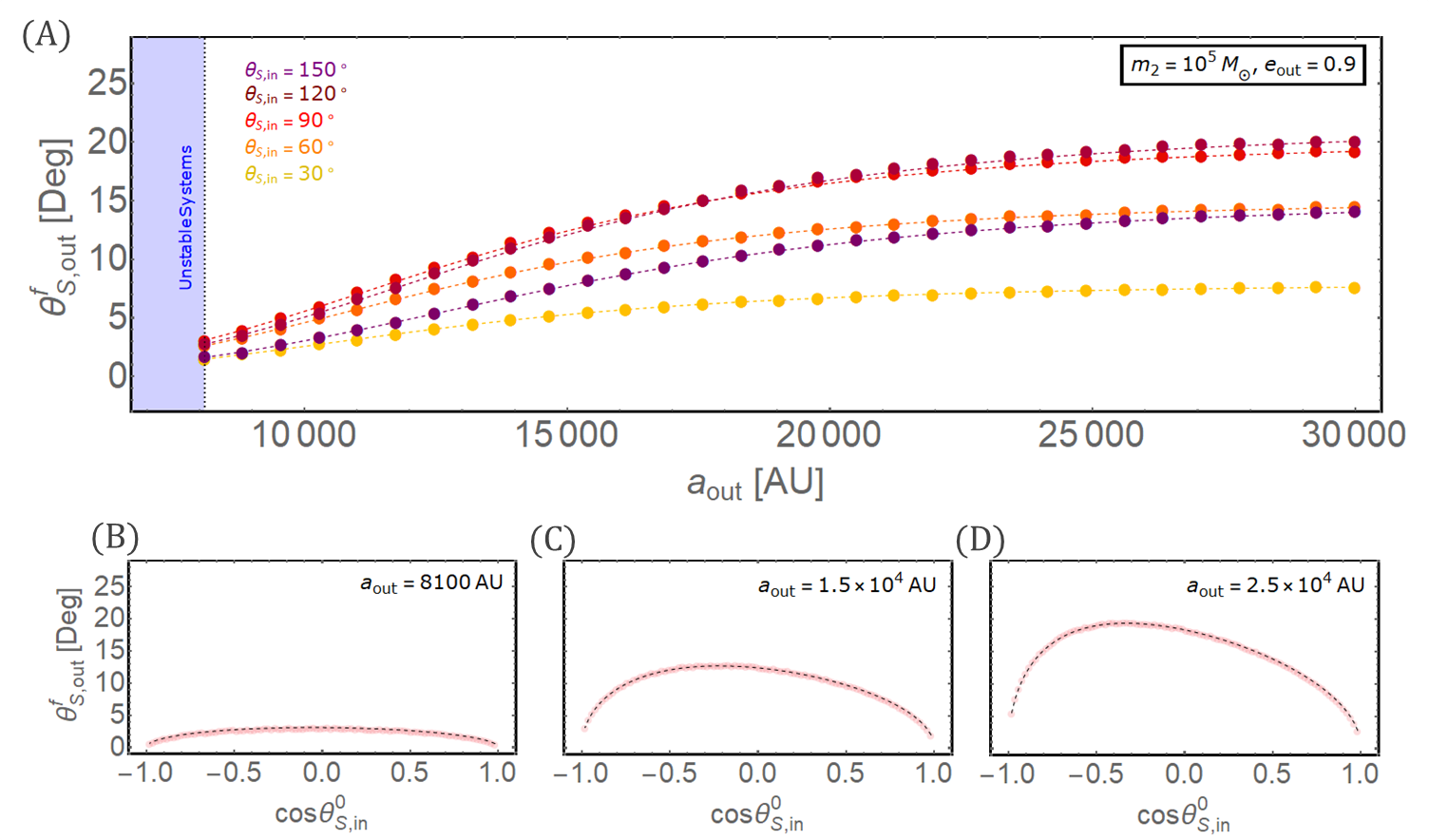}
\caption{Same as Figure \ref{fig:aout dependence 105},
except for $e=0.9$ and larger $a_\OUT$ due to the stability.
}
\label{fig:aout dependence nonzero eout}
\end{centering}
\end{figure*}

Here we consider how the results are changed when the outer orbits have finite eccentricities.

Figure \ref{fig:aout dependence nonzero eout} presents the results from
the fiducial example (see Figure \ref{fig:aout dependence 105}) but with $e=0.9$.
Since the outer eccentricity $e$ only appears in the expression for $\Omega_\OUT^\mathrm{(N)}$,
we carry out the analytical calculations
by using Equation (\ref{eq: Omega OUT Newtonian}) with $e\neq0$.
We find that the numerical results and the analytical calculations
are still in good agreement.

Note that here we do not consider the mutual interactions
between different outer orbits.
For the realistic system, the adjacent eccentric outer orbits
could experience orbital crossings.
But we expect the results for $\theta_{\mathrm{S},\OUT}^\mathrm{f}$
remain largely valid.

\section{Discussion and conclusion}
\label{sec 6}

In this paper, we have studied the secular dynamics
of stars (modeled as test particles)
around a merging massive/supermassive BH binary (BHB),
taking into account the GR effect induced by the rotating BH in the inner binary.
We focus on the circular BHB with relatively small mass ratio,
so that we only need to include the spin of the (more massive) primary BH.
Our goal is to determine the final orbital orientations of the outer
(circumbinary) stellar orbits relative to the spin axis of the merger remanent,
assuming the initial stellar orbital axes are approximately aligned with the
BHB orbital axis.

The evolution of the angular momentum vector of the stellar orbit ($\hat {\bm{l}}$)
is determined by the competition between
the precession of the BHB axis $\hat {\bm{l}}_\IN$
around the primary spin axis $\hat {\bm{S}}_1$ and
the precession of $\hat {\bm{l}}$ around $\hat {\bm{l}}_\IN$.
During the orbital decay of the BHB, the ratio of the two precession rates can
change from $\lesssim1$ to $\gtrsim1$,
leading to a significant change in the orientation of $\hat {\bm{l}}$.
The final direction of $\hat {\bm{l}}$
carries the imprint of the spin of the remanent BH ($\hat {\bm{S}}_1$). Our main findings are:

(\textit{i}) For central BHBs with modest mass ratio ($m_2/m_1\sim0.1$),
there is a quasi-alignment phenomenon for the evolution of the outer stellar orbits.
Namely, starting with nearly coplanar outer orbits (i.e., $\hat {\bm{l}}\parallel\hat {\bm{l}}_\IN$),
the orbital axis $\hat {\bm{l}}$ of the circumbinary star
will preferentially evolve towards the spin direction after the merger of inner BHB,
regardless the initial spin-orbit misalignment angle of the BHB
(see Figure \ref{fig:aout dependence 105}).
This alignment is particularly strong for close stellar orbits.
Such trend of alignment, where the final spin-orbit misalignment angle
($\theta_{\mathrm{S},\OUT}^\mathrm{f}$)
is small, can be understood analytically based
on the principle of adiabatic invariance (Equation \ref{eq: adiabatic invariant}).
Also, our analytical analysis can be applied
to inclined and eccentric outer orbits (Figures \ref{fig:aout dependence inclined Lout 30},
\ref{fig:aout dependence inclined Lout 90} and \ref{fig:aout dependence nonzero eout}).

(\textit{ii}) When the mass ratio of the BHB is more extreme (i.e., $m_2/m_1\lesssim0.01$),
the angular momentum axis of the outer stellar orbit can experience complicated evolution in general.
The adiabaticity condition in the analytical calculation may break down and the
evolution of the stellar orbits can only be resolved numerically by using the full secular equations of motion.
Nevertheless, the alignment effect still works reasonably well when the initial spin-orbit misalignment angle is small
(i.e., $\theta_{\mathrm{S},\IN}^0\lesssim90^\circ$; see Figures \ref{fig:aout dependence 104} and \ref{fig:aout dependence 103}).

There are several caveats in our study:

(\textit{i}) We have neglected the
effect due to the secondary spin in the central BHB. This is
reasonable if the secondary spin $S_2$ is negligible compared to $S_1$ (e.g., when the mass ratio
$m_2/m_1$ is relatively small or when $\chi_2\ll \chi_1$). For comparable-mass BHBs, the final
spin axis the merger remnant is approximately aligned with the
pre-merger orbital axis, thus we expect the circumbinary stellar
orbital axis to be aligned with the final BH spin (assuming
$\hat{\bm{l}}$ is initially aligned with the binary axis).

(\textit{ii}) We have not considered the merger kick acting on the remnant BH, which
may change the orientation of the stellar orbit relative to the final BH spin axis.
For the BHB studied in our paper ($m_1=4\times10^6M_\odot$ and $m_2=10^5M_\odot$, with mass ratio $0.025$),
assuming the primary BH has the maximum spin with isotropic orientation,
the kick velocity ($V_\mathrm{kick}$) on the merger remnant evaluated using the fitting
formula of \citet{Lousto 2010} is less than $\sim40\mathrm{km/s}$.
Compared to the orbital velocity ($V_\mathrm{orb}$) of the stellar orbits studied here ($a_\OUT\lesssim10^5$AU),
we always have $V_\mathrm{orb}\gg V_\mathrm{kick}$. Thus,
the kick effect is negligible.
However, for BHBs with higher mass ratios, the merger kick could play an important role,
especially for the distant stellar orbits with $V_\mathrm{orb}\gtrsim V_\mathrm{kick}$.
In this case, the post-kick orbital orientation can be modified
\citep[e.g.,][]{Liu kick}, and the final spin-orbit misalignment angle must be evaluated based on the
corrected orientation of $\hat{\bm{l}}$. 

(\textit{iii}) We have only considered
BHBs in circular orbit in this paper. When the BHB has a finite eccentricity,
the outer stellar orbit can also gain modest eccentricity through
octupole-order secular interactions \citep[e.g.,][]{Liu et al 2015,Liu APR}.
The finite eccentricity may influence the orbital
inclination evolution indirectly.

Our result suggests that the relative orientation between the spin of
a central massive/supermassive BH and the surrounding stellar orbits might
provide a probe of the merger history of the BH.
In particular, the Galactic Center hosts a population of
young massive stars \citep[e.g.,][]{Ghez 1998,Ghez 2008,Genzel 2000,Merritt 2013,Alexander 2017}.
If the supermassive BH, Sagittarius A$^\ast$,
has experienced a previous merger with an intermediate-mass BH,
it could have left some imprints on the nearby S-star orbits.
It has been suggested that the orbital distribution of S-stars could put constraints on the Sagittarius A$^\ast$ spin
\citep[e.g.,][]{Levin SgrA,Fragione spin}.
Therefore, the precise measurements of the S-star orbits (including the orbital
orientations) and the spin axis of central BH
would be highly desirable.

\section{Acknowledgments}

BL thanks Johan Samsing, Daniel D'Orazio and Adrian Hamers for useful discussion.
DL has been supported in part by NSF grants AST-1715246 and AST-2107796.
This project has received funding from the European Union's Horizon 2020
research and innovation program
under the Marie Sklodowska-Curie grant agreement No. 847523 `INTERACTIONS'.

\section{DATA AVAILABILITY}

The simulation data underlying this article will be shared
on reasonable request to the corresponding author.

\end{document}